\documentclass[12pt]{article}
\usepackage{amsmath}
\usepackage{graphicx}
\usepackage{natbib}
\usepackage{url} 

\newcommand{\blind}{1}

\usepackage{amssymb}
\usepackage{amsmath}
\usepackage{amsthm}
\usepackage{graphicx}
\usepackage{enumerate}

\newtheorem{theorem}{Theorem}[section]

\newtheorem{assumption}{Assumption}[section]

\usepackage{mathtools}
\usepackage{float}
\usepackage{fancyhdr}

\usepackage{booktabs}
\usepackage{bm}
\usepackage{setspace}

\begin{document}

	\def\spacingset#1{\renewcommand{\baselinestretch}%
		{#1}\small\normalsize} \spacingset{1}

	
	\if1\blind
	{
		\title{\bf Double Probability Integral Transform Residuals for Regression Models with Discrete Outcomes}
		\author{Lu Yang\thanks{
				The author gratefully acknowledges 
					the research support from  {the National Science Foundation (DMS-2210712).}}\hspace{.2cm}\\
			School of Statistics, University of Minnesota}
		\date{}
		\maketitle
	} \fi
	
	\if0\blind
	{
		\bigskip
		\bigskip
		\bigskip
		\begin{center}
			{\LARGE\bf Double Probability Integral Transform Residuals for Regression Models with Discrete Outcomes}
		\end{center}
		\medskip
	} \fi
	
\bigskip

\begin{abstract}

		The assessment of regression models with discrete  outcomes is challenging and has many fundamental issues. With discrete outcomes, standard regression model assessment tools such as Pearson and deviance residuals do not follow 
		the conventional reference distribution (normal) under the true model, calling into question the legitimacy of model assessment based on these tools.
		To fill this gap,  
		we construct a new type of residuals  for regression models with general discrete outcomes, including ordinal and count outcomes.
		The proposed residuals are based on two layers of probability integral transformation. 
		When at least one continuous covariate is available,
		the proposed residuals  closely follow a uniform distribution  (or a normal distribution after transformation) under the correctly specified model. One can construct visualizations such as QQ plots to 
		check the overall fit of a model straightforwardly, and
		the shape of  QQ plots can further help identify possible causes of misspecification such as overdispersion. 
				We provide theoretical justification for the proposed residuals by establishing their asymptotic properties.
		Moreover, in order to assess the mean structure and identify potential covariates, we develop an ordered curve  as a supplementary tool, which is based on the comparison between the partial sum of outcomes and of fitted means.
 Through simulation, we demonstrate empirically that the proposed  tools outperform   commonly used residuals  for various model assessment tasks. We also illustrate the workflow of model assessment using the proposed tools in data analysis.

\end{abstract}

\noindent%
\vfill

\newpage
\spacingset{1.9} 

	\section{Introduction}
	Regression models 
	are utilized frequently in many domains of applications to study  relationships between covariates and a discrete outcome of interest. 
	For instance, the effects of treatment on mortality (binary, \citealt{goldman2001effect}), the associations between patients' age and  stage of disease  (ordinal, \citealt{li2012new}), and the relationship between car types  and the number of auto insurance claims (count, \citealt{shi2014multivariate}). When fitting a parametric regression model, model assumptions  including the distribution family and potential covariates  are typically made a priori  based on researchers' knowledge.
	However, researchers’ prior information may not adequately describe the patterns in the data. Resulting model deficiency can lead to biased parameter estimates, misleading conclusions, lack of generalizability of results, and unreliable predictions, among many other detrimental consequences. 
	 Judging a model’s adequacy to describe the data is thus a routine and critical task in statistics. 
	
	Residuals are  regularly employed to assess the agreement between data and an assumed model at hand (\citealt{cook1982residuals}). 	Let $Y_i,~i=1,\ldots,n,$ be the outcome of interest and $\mathbf{X}_i$ be the set of covariates.
	Under a parametric model, which synthesizes information including the distribution family and  regressors, we denote $r(Y_i|\mathbf{X}_i)$ as the generalized model error formulated in \cite{cox1968general}. 
	In a linear regression model, the  independently identically distributed (i.i.d.) errors are given by   $r(Y_i|\mathbf{X}_i)=Y_i-\mathbf{X}_i'\bm{\beta},~i=1,\ldots,n,$ where $\bm{\beta}$ represents the coefficients. 
	\cite{cox1968general} generalized the concept of model errors beyond normality by seeking  i.i.d. unobserved variables. 
	For example, the generalized error for regression models with a continuous outcome, such as a gamma variable, can be  defined as the uniformly distributed  probability integral transform 
	$r(Y_i|\mathbf{X}_i)=F(Y_i|\mathbf{X}_i)$, wherein $F$ is the conditional distribution of $Y_i$ given $\mathbf{X}_i$.
	Generally, when the model is correctly specified, there is  a null distribution $F_r$ such that
	\begin{align}\label{eq:overall}
		\Pr(r(Y_i|\mathbf{X}_i)\leq s)=F_r(s).
	\end{align}
	With the corresponding  parameter estimates plugged in, $\hat{r}(Y_i|\mathbf{X}_i)$ is the residual for the $i$th observation.
	 The distribution of the residuals resembles  that of  the errors. Thus,
	under a correctly specified model,
	the residuals $\hat{r}(Y_i|\mathbf{X}_i),i=1,\ldots,n,$ approximately follow $F_r$. 
	This characteristic leads to a common practice of comparing the empirical distribution of residuals with $F_r$ for the purpose of overall model evaluation.
	Informal graphical tools such as 
	histograms and quantile-quantile (QQ) plots or  formal goodness-of-fit tests such as Kolmogorov–Smirnov and Cramér–von Mises tests to can be employed to assess the closeness between the empirical distribution of residuals and $F_r$.
	Ideally, the null distribution of errors  $F_r$ should be tractable, 
 allowing for convenient assessment.

	The assessment of regression models with discrete outcomes is known to be  challenging due to the lack of tools possessing the desirable attribute \eqref{eq:overall} with a straightforward $F_r$. 
	The  most widely used residuals such as Pearson and deviance residuals  
	do not have a null distribution with a closed form (\citealt{cordeiro2009distribution}).  Typically, they are compared to  a normal distribution.
	However, for discrete outcomes, they
	might not follow a normal distribution even when the model is correctly specified, providing a poor basis for judgment. 
	For the same reason, Cox-Snell residuals (\citealt{cox1968general}), which serve as a compelling diagnostic tool for continuous outcomes, lose effectiveness for discrete outcomes. 

	The difficulties in the assessment of regression models with discrete outcomes   originate from the fact that 
	discrete outcomes  cannot be expressed as transforms of i.i.d. variables whose distributions are free of covariates (\citealt{cox1968general}).
	As a simple example, let $Y_i$ be a binary outcome and $\hat{p}(\mathbf{X}_i)$ be the fitted probability of 1. Its Pearson residual is $\left[Y_i-\hat{p}(\mathbf{X}_i)\right]/\left[\hat{p}(\mathbf{X}_i)(1-\hat{p}(\mathbf{X}_i))\right]^{1/2}$, whose distribution depends on covariates and is not  close to a normal distribution, as conventionally assumed. 
	Without a valid residual distribution  $F_r$  in  \eqref{eq:overall}, 
	assessment routines such as QQ plots 
	become
	questionable.
	
A handful of works are 
	 devoted  to   model assessment  for discrete outcomes.
	\cite{anscombe1953contribution} and \cite{pierce1986residuals} proposed the approach of  
	 creating approximately i.i.d. variables. 
	 However,
	 this approximation    might be unsatisfactory, especially for highly discrete data such as binary outcomes. 
	  \cite{ben2004quantile} adhered to deviance residuals and 
	 proposed to  estimate the distributions of deviance residuals beforehand in order to construct a QQ plot, while \cite{davison1989deviance} proposed to use a normal plot.
	In order to retain the favorable properties of residuals for continuous outcomes,
	randomized quantile residuals and their variations (\citealt{dunn1996randomized,liu2018residuals}) are based on the idea of
	 filling the gaps in discrete outcomes using a simulated continuous random variable.
	 An artificial layer of noise is introduced to the data by the nature of this expedient.
	 \cite{shepherd2016probability} developed  residuals applicable to different types of data, although they  do not follow the null pattern (uniformity) under discreteness.
	\cite{yang2021assessment} proposed an alternative to the empirical distribution  of Cox-Snell residuals for discrete data. It can keep the null pattern when the model is correctly  specified and deviates from the pattern when the model is misspecified. Nonetheless, as the output is a function rather than residuals, this tool does not provide clues on what could possibly go wrong. 
There are also tools built for specific types of discrete data, for instance binary (\citealt{landwehr1984graphical,miller1991validation}) and more general ordinal data  (\citealt{liu2018residuals}). 

	In this paper, we construct a new type of residuals for discrete outcomes in Section \ref{sec:method}. 
	We consider  settings where at least one continuous covariate is available and thus no grouped data are available, 
	rendering conventional tools such as deviance residuals unhelpful.
	The proposed residuals are based on two layers of probability integral transformation. The associated errors  are  i.i.d. following a uniform distribution under the null, which is  tractable and achieves \eqref{eq:overall}. 
	One can also conduct a normal quantile transformation, and  the normality is the null pattern.
	As a result, we can construct visualizations such as QQ plots to 
	check the overall model fit straightforwardly.
	We further provide insights on the shape of  QQ plots, which can help identify potential causes of misspecification.
					We provide theoretical justification for the proposed residuals by establishing their asymptotic properties in Section \ref{sec:asym}.
	Moreover, to assess the mean structures specifically, in Section \ref{sec:meancurve},  we propose an ordered curve which is based on the comparison between the partial sum of outcomes and of fitted means. The ordered curve can help detect inadequacies in the mean structure. 
	Combining the ordered curve with the proposed residuals, one can  find  directions for model improvement precisely.

To get a better understanding of situations in which the proposed tools can be useful, Section \ref{sec:simulation} provides a detailed simulation study. Additionally, we  present real data  applications and illustrate the workflow of model assessment using the proposed tools  in Section \ref{sec:data}. Concluding remarks are provided in Section \ref{sec:conc}. Longer proofs of theorems 
are in the supplementary material.
	
	\section{Double Probability Integral Transform Residuals}\label{sec:method}
		Let $Y$ be the discrete outcome of interest.  Without loss of generality, we assume that $Y$  can only take nonnegative integers.
		Denote the underlying distribution function of $Y$ conditional on  covariates
		$\mathbf{X}=\mathbf{x}$ as $F(y|\mathbf{x})=\Pr(Y\leq y|\mathbf{X}= \mathbf{x})$.    
		Under a parametric model $M$, we denote   ${F}_M$ as the corresponding distribution of $Y$ given $\mathbf{X}$.
		
		\subsection{Cox-Snell Residuals for Continuous Outcomes}
		
		 Plugging $(\mathbf{X},Y)$ in $F$, the variable $F(Y|\mathbf{X})$  is known as the probability integral transform. 
		 If $Y$ is a continuous outcome,
		for any fixed value $s \in (0, 1)$, 
		\begin{align*}
			\Pr\left(F(Y|\mathbf{X})\leq s|\mathbf{X}= \mathbf{x}\right)=s.
		\end{align*}
		Taking the expectation with respect to $\mathbf{X}$ yields 
		$	\Pr(F(Y|\mathbf{X})\leq s)=s.$
		That is,
		$F(Y|\mathbf{X})$ is uniformly distributed for continuous $Y$. 
		We can see that the property \eqref{eq:overall} 
		holds for the   probability integral transform of continuous outcomes.
		
\begin{figure}[!h]
			\centering
			\includegraphics[width=.6\textwidth]{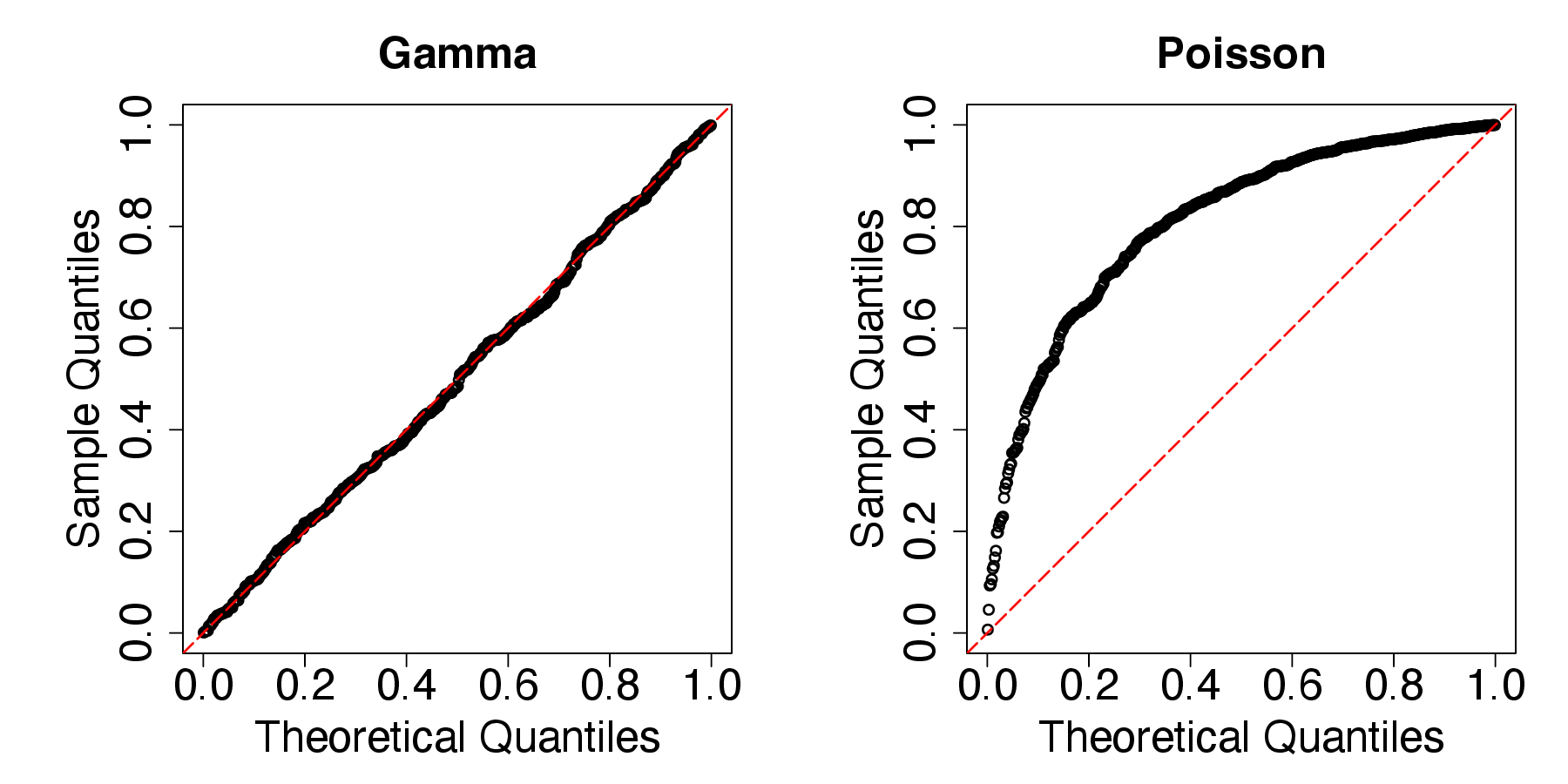} 
			
			\caption{QQ plots of Cox-Snell residuals under correctly specified models. Left panel: gamma regression. 
				Right panel:  Poisson regression. The dashed line is the diagonal throughout the paper. \label{fig:curves}}
		\end{figure}

		Given an i.i.d. sample  $(\mathbf{X}_{i},Y_{i}), i = 1, \ldots , n$, one can acquire the fitted distribution  $\hat{F}_M$  using  parameter estimates.
		Then  a sequence of Cox-Snell residuals, $\hat{F}_M(Y_{i}|\mathbf{X}_{i}),i = 1, \ldots , n$, 
		can be calculated. 
		If the model is correctly specified, 
		the Cox-Snell residuals should be approximately uniformly distributed,
		and  an otherwise large discrepancy with uniformity indicates misspecification. 
		Figure \ref{fig:curves} portrays the  distribution  of the Cox-Snell residuals in simulated examples.
		In the left panel, the data are generated with a gamma regression model, and the Cox-Snell residuals are obtained under the correct model. As anticipated, the Cox-Snell residuals appear to be uniformly distributed. 

		However, if $Y$ is a discrete variable, 
		the Cox-Snell residuals may not be uniformly distributed even under the true model. 
		In the right panel of  Figure \ref{fig:curves}, the data are generated from a Poisson generalized linear model (GLM). It displays the QQ plot of the Cox-Snell residuals when the model is correctly specified, which nevertheless shows a substantial deviation from the diagonal line.

		\subsection{Proposed Residuals}
				Our goal is to construct a new type of residuals for discrete outcomes which satisfies the desirable property  \eqref{eq:overall} with a convenient  null distribution $F_r$.
		We first study the
		probability integral transform of  discrete outcomes. 
		Given $\mathbf{X}=\mathbf{x}$,
		\begin{align}\label{eq:condi}
			\Pr\left(\left.F(Y|\mathbf{X})\leq s\right|\mathbf{X}=\mathbf{x}\right)=	\Pr\left(\left.Y\leq F^{(-1)}(s|\mathbf{x})\right|\mathbf{X}=\mathbf{x}\right)=F\left( \left.F^{(-1)}(s|\mathbf{x})\right|\mathbf{x}\right),
		\end{align}
		where $F^{(-1)}(s|\mathbf{x})\coloneqq\sup\{k\in \mathbb{N}:F(k|\mathbf{x})\leq s\}$. 
		Note that $F^{(-1)}$ differs from the commonly used  definition of 
		the inverse cumulative distribution function.  
		For completeness, we let $F^{(-1)}(0|\mathbf{x})=-\infty$. 
		Taking  the expectation over $\mathbf{X}$, it follows that
		\begin{align}\label{eq:theom}
			G_{0}(s)\coloneqq\Pr\left(F(Y|\mathbf{X})\leq s\right)
			=\mathrm{E}_\mathbf{X}\left[\left.F\left( F^{(-1)}(s|\mathbf{X})\right|\mathbf{X}\right)\right].
		\end{align}
		We can see that, first, 
when $Y$ is discrete,  the composition  $F\left( \left.F^{(-1)}(s|\mathbf{x})\right|\mathbf{x}\right)$ is not necessarily equal to $s$. Consequently,
		$G_{0}$ in \eqref{eq:theom} is not the identity function, and thus the probability integral transform is not uniformly distributed, as illustrated in the right panel of Figure \ref{fig:curves}.  
		Second, if 	 $Y$ is continuous, \eqref{eq:theom} simplifies to the identity function. 

		If $\mathbf{X}$ contains continuous components, 
		 $F(Y|\mathbf{X})$ becomes a continuous variable as a transform of $\mathbf{X}$ and $Y$; see details in the supplementary material. 
		 Although  $F(Y|\mathbf{X})$ itself is not uniformly distributed for discrete outcomes, 
		 because $F(Y|\mathbf{X})$ is
		 a continuous variable,
		 another layer of probability integral transformation, namely $G_{0}\left(F(Y|\mathbf{X})\right)$, produces a uniform variable under the true model. We call 
		$	{r}(Y|\mathbf{X})=G_{0}\left(F(Y|\mathbf{X})\right)$  the \textit{double probability integral transform}  (DPIT for brevity), whose distribution $F_r$ is  uniform and straightforward to work with. 	The DPIT serves as our generalized error in \eqref{eq:overall}. A list of the DPIT properties is provided in Appendix \ref{sec:add}.

		With a sample $(\mathbf{X}_i,Y_i),~i=1,\ldots,n$, 
		now we construct residuals based on the DPIT.
	With a model $M$,
		 one could theoretically use ${G}_{0}\left(\hat{F}_M(Y_{i}| \mathbf{X}_{i})\right)$, $i=1,\ldots,n$. 
		The distribution of the probability integral transform  for a discrete outcome, $G_{0}$,  yet remains to be specified and estimated in practice.
	
	\sloppy
		Based on \eqref{eq:theom}, an intuitive estimator for ${G}_{0}(s)$ is the empirical mean
		$\frac{1}{n}\sum_{j=1}^n {F}\left(\left.{F}^{(-1)}(s| \mathbf{X}_{j})\right| \mathbf{X}_{j}\right).$
		There are two issues  to be addressed here. First, this raw estimator depends on $F$, which  is unknown in practice. 
		We use the model-based version  with estimated parameters plugged in, $\hat{F}_M$.
		  Second, 
		  when constructing the residual for the $i$th observation, we should  estimate ${G}_{0}$ 
		  using other independent observations to avoid bias; see elaboration in Appendix \ref{sec:add}.
Taken together, we  develop an estimator of $G_{0}$ suited to the $i$th observation
		\begin{align}
			\label{eq:empgm}
			\hat{G}_{Mi}(s)=	\frac{1}{n-1}\sum_{j=1,j\neq i}^n \hat{F}_M\left(\left.\hat{F}_M^{(-1)}(s| \mathbf{X}_{j})\right| \mathbf{X}_{j}\right).
		\end{align}
%
While $\hat{G}_{Mi}$ is a function defined for $s\in (0,1)$, we  only need its value at one point,  $\hat{F}_M(Y_{i}| \mathbf{X}_{i})$, to develop the residuals.
	
\begin{figure}[!h]
			\centering
			\includegraphics[width=.6\textwidth]{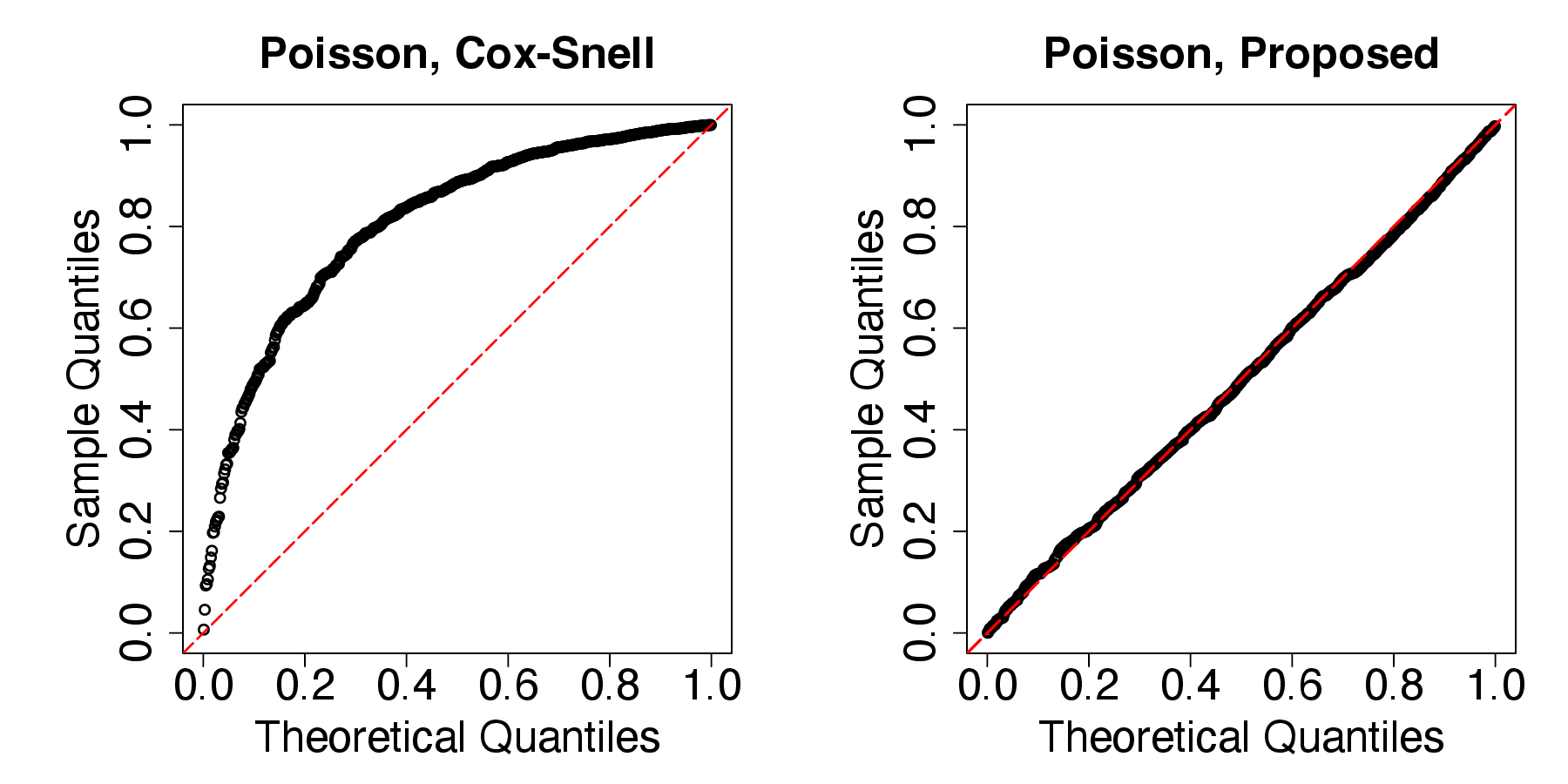} 
			
			\caption{QQ plots of  residuals  under the correctly specified model. \label{fig:curvespois}}
		\end{figure}

	Combining the components above, we propose the \textit{double probability integral transform residuals} (DPIT residuals for brevity)
\begin{align}\label{eq:propose}
	\hat{r}(Y_i|\mathbf{X}_i)=\hat{G}_{Mi}\left(\hat{F}_M(Y_{i}| \mathbf{X}_{i})\right),  i = 1, \ldots , n .
\end{align}
		{If the model is correctly specified,  the DPIT residuals   should closely follow a uniform  distribution (e.g., the right panel of Figure \ref{fig:curvespois}), and otherwise model deficiency is implied}.
				To facilitate visualization and  comparison with other residuals,
		one can also  apply the    normal  quantile transformation $\Phi^{-1}$  to the DPIT residuals, 
		resulting in $$\Phi^{-1}\left[	\hat{r}(Y_i|\mathbf{X}_i)\right],i=1,\ldots,n.$$
		Consequently,   a standard  normal distribution serves as the  null pattern. In Figure \ref{fig:normal} of Appendix \ref{sec:addsimu}, we demonstrate the behavior of the residuals on both uniform and normal scales in a simulated example, wherein we can see that  the uniform residuals accentuate the center values, while the normal residuals emphasize  the tails.
		
			The construction of the proposed residuals involves two-stage estimation. We need to first estimate the model parameters using all the data to obtain $\hat{F}_M$. Subsequently, using the parameter estimates, we estimate $G_{0}$ through $\hat{G}_{Mi}$ for each observation $i$.
					 	For the $i$th observation, 
					 	the corresponding residual can be calculated as follows.
		\begin{enumerate}
			\item Calculate $a_i=\hat{F}_M(Y_i|\mathbf{X}_i)$.
			\item For $j= 1,\ldots,n$ and $j\neq i$, calculate $b_{ij}=\hat{F}_M^{(-1)}\left(a_i|\mathbf{X}_j\right)$, and then determine $c_{ij}=\hat{F}_M\left(b_{ij}|\mathbf{x}_j\right)$.
			\item Compute the average of $c_{ij}$ to yield
			$\hat{r}(Y_i|\mathbf{X}_i)=\frac{1}{n-1}\sum_{j=1,j\neq i}c_{ij}.$
			 To acquire the residuals on the normal scale, one can further apply  $\Phi^{-1}\left[	\hat{r}(Y_i|\mathbf{X}_i)\right].$
		\end{enumerate}

		We construct  the DPIT residuals by  compressing the Cox-Snell residuals using $\hat{G}_{Mi}$, 
		 thereby ensuring the residuals' uniform distribution for discrete outcomes under the null. 
		Since $\hat{G}_{Mi}$ converges to $G_0$ which is a monotone function, the proposed residuals nearly preserve the ordering of  the Cox-Snell residuals.
		As illustrated in Figure \ref{fig:curvespois}, 
		$\hat{G}_{Mi}$ brings the Cox-Snell residuals displayed in the left panel to the diagonal in the right panel. 
		Lastly,
			if $Y$ is continuous, \eqref{eq:empgm} simplifies to the identity function, and thus the proposed residuals coincide with  Cox-Snell residuals. Hence, Cox-Snell residuals for continuous data can be viewed as a special case of the proposed residuals.
		

			\subsection{Handling  Ordinal and Binary Outcomes}
\begin{figure}[!h]
			\centering
			\includegraphics[width=.9\textwidth]{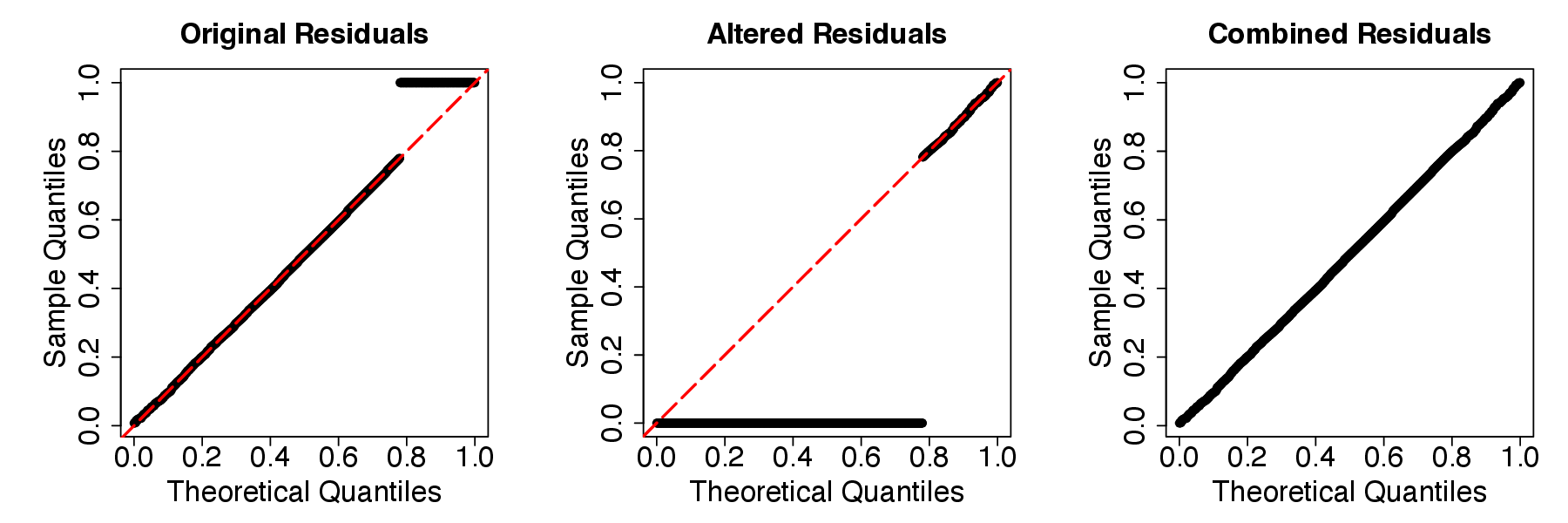} 
			
			\caption{QQ plots of the proposed residuals for binary outcomes with a correctly specified model.\label{fig:binary}}
		\end{figure}

		For ordinal outcomes including binary, the probability integral transform $F(Y|\mathbf{X})$ is situated at 1 if $Y$ takes the largest possible value. %
		Suppose the possible values of $Y$ are $0,1,\ldots,k_{\max}$, 
		then $F(Y|\mathbf{X})$ has a point mass at 1 with probability
		$$\Pr\left(F(Y|\mathbf{X})=1\right)=\Pr\left(Y=k_{\max}\right)=1-\mathrm{E}_{\mathbf{X}}F({k_{\max}-1}|\mathbf{X}).$$
		The distribution of the DPIT is 
		\begin{align*}
			\Pr\left(G_{0}(F(Y|\mathbf{X}))\leq s\right)=\begin{cases*}
				1&$s=1$,\\
				\mathrm{E}_{\mathbf{X}}F({k_{\max}-1}|\mathbf{X})&$\mathrm{E}_{\mathbf{X}}F({k_{\max}-1}|\mathbf{X})<s< 1,$\\
				s&$s\leq \mathrm{E}_{\mathbf{X}}F({k_{\max}-1}|\mathbf{X})$.
			\end{cases*}
		\end{align*}
	See derivations in the supplementary material. 
		We can see from the equation above that for  $\mathrm{E}_{\mathbf{X}}F({k_{\max}-1}|\mathbf{X})<s<1,$ 
	$\Pr\left(G_{0}(F(Y|\mathbf{X}))\leq s\right)$ does not vary with $s$.
	Consequently, when constructing QQ plots,  the  quantiles of $G_{0}(F(Y|\mathbf{X}))$  with a probability $s>\mathrm{E}_{\mathbf{X}}F({k_{\max}-1}|\mathbf{X})$ clump at 1.
		Hence, 
		this area is not helpful for model assessment. 
		As demonstrated in the left panel of  Figure \ref{fig:binary}, in which the simulated outcomes are binary and the model is correctly specified, a cluster of points appears at 1. The points corresponding  to the observations whose response variable takes value at 0, in contrast, closely align with the diagonal  and provide a correct signal. 
		
%

To make full use of the data, we consider an altered 
DPIT, 
${H}_{0}\left({F}(Y-k_{\max}|\mathbf{X}_i)\right)$, wherein $H_{0}(s)\coloneqq\Pr\left(F(Y-k_{\max}|\mathbf{X})\leq s\right)
=\mathrm{E}_\mathbf{X}\left[F\left( F^{(-1)}(s|\mathbf{X})+k_{\max}|\mathbf{X}\right)\right].$ 
The corresponding altered 
residuals are \begin{align}\label{eq:alter}
	\hat{r}_{S}(Y_i|\mathbf{X}_i)=\hat{H}_{Mi}\left(\hat{F}_M(Y_i-k_{\max}|\mathbf{X}_i)\right)
\end{align}
where  $$\hat{H}_{Mi}(s)=\frac{1}{n-1}\sum_{j=1,j\neq i}^n \hat{F}_M\left(\hat{F}_M^{(-1)}(s| \mathbf{X}_{j})+k_{\max}| \mathbf{X}_{j}\right).$$
In the middle panel of Figure \ref{fig:binary}, we display the altered residuals, which exploit the information of the data  with outcomes equal to 1.
The observations whose responses are  0, on the other hand, do not provide information in this case. 

To integrate, we  combine both types of  residuals for ordinal outcomes.
One should use the original DPIT residuals for  observations  smaller than the largest possible value and employ the  altered residuals for  observations taking value at the maximum. 
By doing so, we obtain residuals whose null distribution is uniform on $(0,1)$. In the right panel of Figure \ref{fig:binary}, the residual plot utilizes all the data and is close to the diagonal with  a correctly specified model.

			\subsection{Large Sample Properties}\label{sec:asym}
			
		In this section, we study the asymptotic distribution of the proposed residuals. 
		In particular, we look into the empirical residual distribution function 
		\begin{align*}
\frac{1}{n}\sum_{i=1}^n1\left[\hat{r}(Y_i|\mathbf{X}_i)\leq s\right]
				=\frac{1}{{n}}\sum_{i=1}^n 1\left[\hat{G}_{Mi}\left(\hat{F}_M(Y_{i}| \mathbf{X}_{i})\right)\leq s\right].
		\end{align*}
			We focus on the null behavior of the proposed tool,
		 namely when the model is correctly specified $F=F_M$.
	For clarity, we index  functions by their parameters $\bm{\beta}$ in this section. The underlying parameters $\bm{\beta}_0$ are estimated using the maximum likelihood estimator $\hat{  \bm\beta}$.
		The  quantity of interest to study is then
		\begin{align}\label{eq:xn}
\frac{1}{\sqrt{n}}\sum_{i=1}^n \left\lbrace1\left[\hat{G}_{\hat{  \bm\beta}i}\left({F}(Y_{i}| \mathbf{X}_{i},\hat{  \bm\beta})\right)\leq s\right]-s\right\rbrace.
		\end{align}
where
$$	\hat{G}_{\hat{  \bm\beta}i}(s)=	\frac{1}{n-1}\sum_{j=1,j\neq i}^n {F}\left(\left.{F}^{(-1)}(s| \mathbf{X}_{j},\hat{  \bm\beta})\right| \mathbf{X}_{j},\hat{  \bm\beta}\right).$$
	\begin{theorem}\label{main}
		
		Under   Assumptions \ref{op}, \ref{bound}, and \ref{lips}, 
		when the model is correctly specified, 
		  for fixed $s$, $
		  \frac{1}{\sqrt{n}}\sum_{i=1}^n \left\lbrace1\left[\hat{G}_{\hat{  \bm\beta}i}\left({F}\left(Y_{i}| \mathbf{X}_{i},\hat{ \bm \beta}\right)\right)\leq s\right]-s\right\rbrace
		  $  converges weakly
		to a Gaussian distribution. Its  mean is $\mathrm{E}g_s(\mathbf{X},Y)$  and variance is $\mathrm{E}g_s(\mathbf{X},Y)^2-\left(\mathrm{E}g_s(\mathbf{X},Y)\right)^2$, where 
		$$g_{s}(\mathbf{x},y)=
		f_{s}(\mathbf{x},y)+\left[I(\bm\beta_0)\right]^{-1}\left.\frac{\partial \left[\tilde{G}_{\bm\beta,\bm{\beta}_0}\left(G_{\bm\beta}^{-1}(s)\right)\right] }{\partial \beta}\right\vert_{\beta=\bm\beta_0}\dot{l}_{\bm\beta_{0}}(\mathbf{x},y),$$
		and	
$$f_s(\mathbf{x},y)=1\left(G_0(F(y|\mathbf{x},\bm{\beta}_0))\leq s\right)-F\left(\left.F^{(-1)}(G_0^{-1}(s)|\mathbf{x},\bm{\beta}_0)\right|\mathbf{x},\bm{\beta}_0\right).$$
Furthermore,
\begin{align}\label{eq:tildeg}
\tilde{G}_{\bm\beta,\bm{\beta}_0}(s)\coloneqq\Pr\left(F(Y|\mathbf{X},\bm{\beta})\leq s\right)=\mathrm{E}_\mathbf{X}\left[F\left(\left. F^{(-1)}(s|\mathbf{X},\bm\beta)\right|\mathbf{X},\bm\beta_0\right)\right];
\end{align}
 $I(\beta)$ is the Fisher information matrix; ${l}_{\beta}(\mathbf{x},y)$ is   the log-likelihood, and $\dot{l}_{\bm\beta}(\mathbf{x},y)=\partial  l_{\bm\beta}(\mathbf{x},y)/\partial\bm\beta$ is the score function. 
	\end{theorem}

From Theorem \ref{main}, we can see that the proposed residuals do converge to being uniformly distributed asymptotically, at the order of $\sqrt{n}$.
Furthermore, the uncertainties associated with the distribution of the proposed residuals originate from two distinct sources.
The first part $f_s(\mathbf{x},y)$ arises from the estimation of $G_0$, and the second part involves the uncertainty in $\hat{  \bm\beta}$.
	The details and proof of the theorem are provided in the supplementary material.
	
			\section{Ordered  Curve}\label{sec:meancurve}
			Reflecting on \eqref{eq:condi}, the distribution of the probability integral transform varies with the value of $\mathbf{x}$ for discrete outcomes. When constructing the proposed residuals, despite an overall transformation $G_0$, the dependence of the DPIT on  covariates remains. As will be demonstrated in the simulation study, the residuals versus predictor plots are not informative. 
			Therefore, the DPIT residuals  have limited capacity of detecting deficiency  in the mean structure. 
			In this section, we propose a supplementary tool for  assessing   mean structures. 
		
		Lorenz curves  (\citealt{hand1997statistical}), a concept originally employed to compare  risk classifiers in finance, were adapted by \cite{frees2011summarizing} to judge the adequacy of insurance premiums. 
		In their framework, Lorenz curves compare the cumulative sum of  premiums with the cumulative sum of  actual losses, as a threshold changes. 
		Here we generalize this idea to assess   the mean structure of discrete outcomes.
	
		In order to assess the mean structure, we propose to compare the cumulative sum of the response variable $Y$ and its hypothesized value.
		Denote the mean for the $i$th observation as $\lambda_i$, and $\hat{\lambda}_i$ is the fitted value.  For instance, in a Poisson GLM with a log link, $\lambda_i=\exp\left(\mathbf{X}_i'\bm\beta\right)$. 
				Note that for ordinal outcomes, it requires   the assignment of  numbers to categories, and $\lambda_i$ is the mean  based on the  integer encoding of categories. 
		We further let $\Lambda$, which is a transform of $\mathbf{X}$, be the  random variable generating $\lambda_i$. 
		We introduce a  variable denoted as $Z$ to determine the cutoff. Here $Z$ can be $\Lambda$ itself, a linear combination  of  $\mathbf{X}$,  or a variable  outside  $\mathbf{X}$. We will discuss below the choice of $Z$. 
		On the one hand, the cumulative response function is defined as $$L_1(t)=\frac{\mathrm{E}\left[Y1(Z\leq t)\right]}{\mathrm{E}\left[Y\right]},$$
		with its empirical counterpart being $$\hat{L}_1(t)=\frac{\sum_{i=1}^n\left[Y_i1(Z_i\leq t)\right]}{\sum_{i=1}^nY_i}.$$
		The function is divided by the sum of the responses for normalization.
		On the other hand, assuming the model is correct,
	if $Z$ is either contained in $\mathbf{X}$ or independent of $\mathbf{X}$ and $Y$, 
	$$\mathrm{E}\left[Y1(Z\leq t)\right]
	=\mathrm{E}\left[\Lambda1(Z\leq t)\right];$$ see Appendix \ref{sec:add} for details.
		In either case, we define $$L_2(t)=\frac{\mathrm{E}\left[\Lambda1(Z\leq t)\right]}{\mathrm{E}\Lambda}.$$
		Its empirical version is $$\hat{L}_2(t)=\frac{\sum_{i=1}^n\left[\hat{\lambda}_i1(Z_i\leq t)\right]}{\sum_{i=1}^n\hat{\lambda}_i}.$$
		If the mean structure is correctly specified, namely $\Lambda$ is indeed the mean of $Y$, we expect that $L_1$ and $L_2$ are close to each other.
		Therefore, in practice,  we can examine the  curve $(\hat{L}_2(t),\hat{L}_1(t))$, as $t$ varies. 
		If the  curve is distant from the diagonal, it suggests
		discrepancies between $L_1$ and $L_2$ and thus  
		 incorrectness in the mean structure.
				In addition, the  position of the curve relative to the diagonal reveals 
		the  relationship between the fitted and underlying means. 
		If the curve lies above the diagonal,  it indicates that the cumulative sum  of the response is larger than that of the fitted mean, 
		and thus  the mean  is underestimated, and vice versa.

		This idea of comparing the estimated  mean to the   mean of the outcome is  used to construct  calibration plots for classifiers.
		   To do so, the fitted probabilities are typically binned (\citealt{faraway2016extending}).
		 To   avoid the ad-hoc nature of binning,
		 we consider the cumulative sum instead.
		    \cite{su1991lack} used the  partial sums of residuals over a  partition of the covariate space. However,  their method required simulations to establish the null behavior of the tool.
		
		To pragmatically construct the ordered curve,
		one can order the threshold variable $Z_{(1)},Z_{(2)},\ldots,Z_{(n)}$ and denote the original index  of $Z_{(i)}$  as $T_{i}$. That is, $Z_{(i)}=Z_{T_i}$.
		The empirical version of the curve  $\hat{L}_1(t)$ against $\hat{L}_2(t)$  can then be characterized by the following $n$ points
		\begin{align*}
			 \left(\frac{\hat{\lambda}_{T_1}}{\sum_{i=1}^n\hat{\lambda}_{i}},\frac{Y_{T_1}}{\sum_{i=1}^nY_{i}}\right),\left(\frac{\hat{\lambda}_{T_1}+\hat{\lambda}_{T_2}}{\sum_{i=1}^n\hat{\lambda}_{i}},\frac{Y_{T_1}+Y_{T_2}}{\sum_{i=1}^nY_{i}}\right),\ldots,\left(\frac{\sum_{i=1}^k\hat{\lambda}_{T_i}}{\sum_{i=1}^n\hat{\lambda}_{i}},\frac{\sum_{i=1}^kY_{T_i}}{\sum_{i=1}^nY_{i}}\right),\ldots,\left(1,1\right).
		\end{align*} 
	In GLMs,
		as a result of the normal equation,   $\sum_{i=1}^n\hat{\lambda}_{i}$ and $\sum_{i=1}^nY_{i}$ are typically very close. Hence, the denominators of $\hat{L}_1$ and $\hat{L}_2$ serve as normalization factors and do not impact the shape of the curve much.

		The role of the threshold variable $Z$ is to determine the rule  for accumulating $\Lambda$ and $Y$ for the ordered curve. The candidates for $Z$ include first the  fitted values, second a linear combination of $\mathbf{X}$, or third a potential variable to be included as a covariate.
In the third case, 
if the  variable being considered is in fact irrelevant, it is equivalent to randomly reordering the data and calculating the partial sums.

		To reveal a potential lack of fit, the optimal choice for  $Z$ should lead to a significant separation between $\Lambda$ and $Y$. Assuming no collinearity, an  important  predictor which is missing in the model  can induce such an effect, since it is highly correlated with $Y$ but not with $\Lambda $. 
Therefore, if a variable  leads to a large discrepancy between the ordered curve and the diagonal, including this variable in the mean function should be considered.

		\section{Simulation}\label{sec:simulation}
		In this section, we discuss the operating characteristics of the proposed tools  via various examples. In Section \ref{sec:correct}, we show that the proposed residuals exhibit null patterns when the model is correctly specified. In Section \ref{sec:miss}, we explore their behaviors when the model is misspecified. In Section \ref{sec:simmean},  the empirical performance of the ordered  curve is evaluated.
		\subsection{Closeness to Null Pattern under True Models}\label{sec:correct}
		
\begin{figure}[!h]
			\centering
			\includegraphics[width=.9\textwidth]{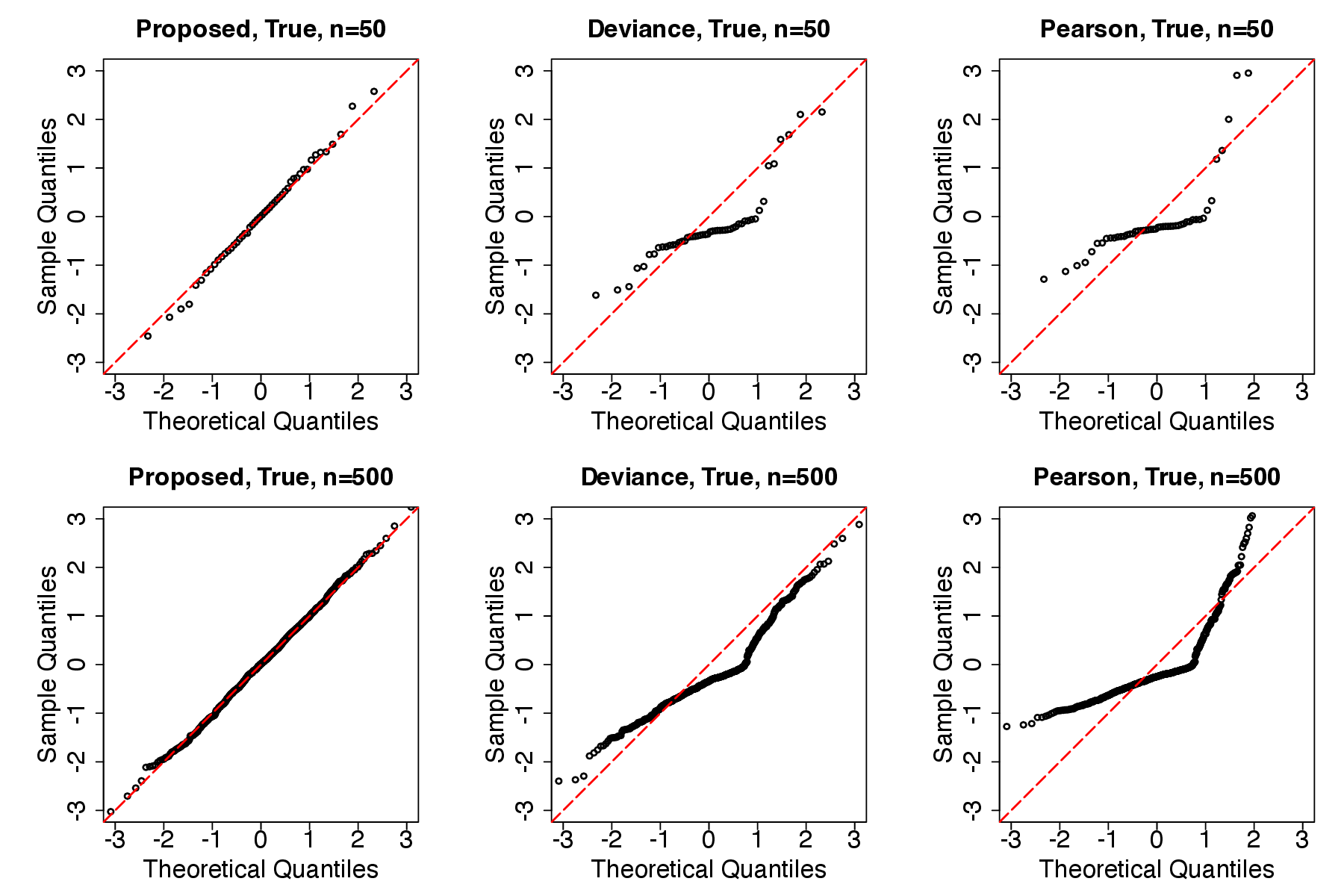} 
			
			\caption{
				 QQ plots of the proposed residuals (left column)  compared with deviance (middle column) and  Pearson residuals (right column) when the model is correctly specified. The sample size is 50 in the top row and 500 in the second row. \label{fig:true}}
		\end{figure}

		We first demonstrate that when the model is correctly specified, the proposed residuals follow the null pattern. 
		In Figure \ref{fig:true}, the data are generated using a negative binomial distribution with  mean   $\mu =  \exp\left(\beta_0 +X_1\beta_1+X_2\beta_2\right)$, where $X_1 \sim N(0,1)$, and $X_2$ is binary with a probability of success as 0.7. The coefficients  are set as $\beta_0 = -2, \beta_1 = 2$, and  $\beta_2=1$. The underlying size parameter is 2. We vary the sample size  from 50 to 500.
For visual clarity and comparability with other residuals,
	 we  present our residuals on the normal scale $\Phi^{-1}\left[\hat{r}(Y_i|\mathbf{X}_i)\right]$ in this section, 
		and thus  a standard  normal distribution is the  null pattern.
		
		In the left column of Figure \ref{fig:true}, when the model is correctly specified, the QQ plots of the proposed residuals  are in close  proximity  to the diagonal (dashed line throughout this section), indicating  that they closely follow the null distribution. This behavior persists even when the sample size is rather  small in the top row. This is an advantage of the proposed residuals over the tool in \cite{yang2021assessment}, which utilizes kernel functions and requires a large sample size.
		For comparison, we also display the normal QQ plots of   deviance and Pearson residuals. They deviate from the null pattern and erroneously signal a deficiency in the model, leading to a Type I error.

\begin{figure}[!h]
		\centering
		\includegraphics[width=.9\textwidth]{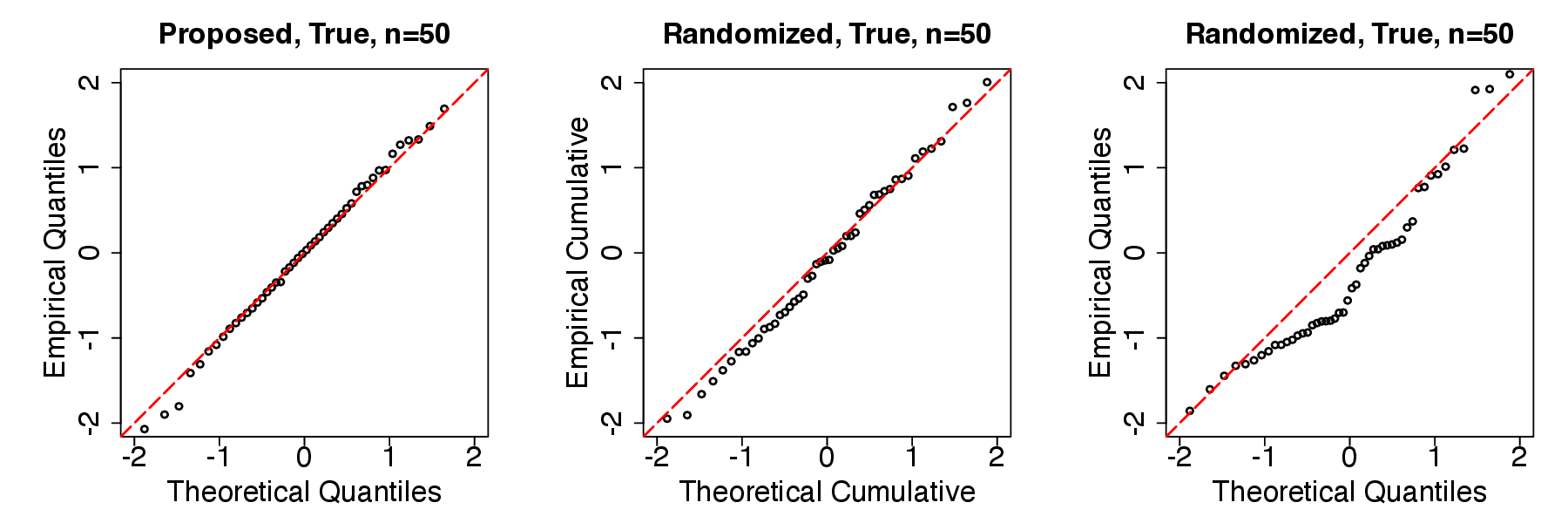} 
		
		\caption{QQ plots of the proposed residuals (left) and randomized quantile residuals (middle and right panels, with different seed numbers) for negative binomial outcomes with a correctly specified model.\label{fig:seed}}
	\end{figure}

	We further compare the proposed residuals with the randomized quantile residuals in Figure \ref{fig:seed}. The randomness associated with randomized quantile residuals  is particularly evident for small sample sizes. For the same dataset and  model, we see distinct patterns in the randomized quantile residuals with different seed numbers,  as shown  in the middle and right panels. 
	
		To assess the variability  of the proposed residuals, we further present a simulation study with 10000 replicates  in the supplementary material. It shows that our residuals are close to being normally distributed 
		under small sample sizes.
		To save space,  the null patterns of the proposed residuals under other distributions are displayed in the comparative plots of Section \ref{sec:miss}.
		
	\subsection{Discrepancy with the Null When the Model Is Misspecified}\label{sec:miss}
	In this section, we demonstrate that the proposed residuals can effectively detect model misspecification in various scenarios. 
	Sample sizes are 500 throughout this section.

		\subsubsection{Overdispersion and Zero-Inflation in Count Data}\label{sec:count}
\begin{figure}[!h]
	\centering
	\includegraphics[width=.9\textwidth]{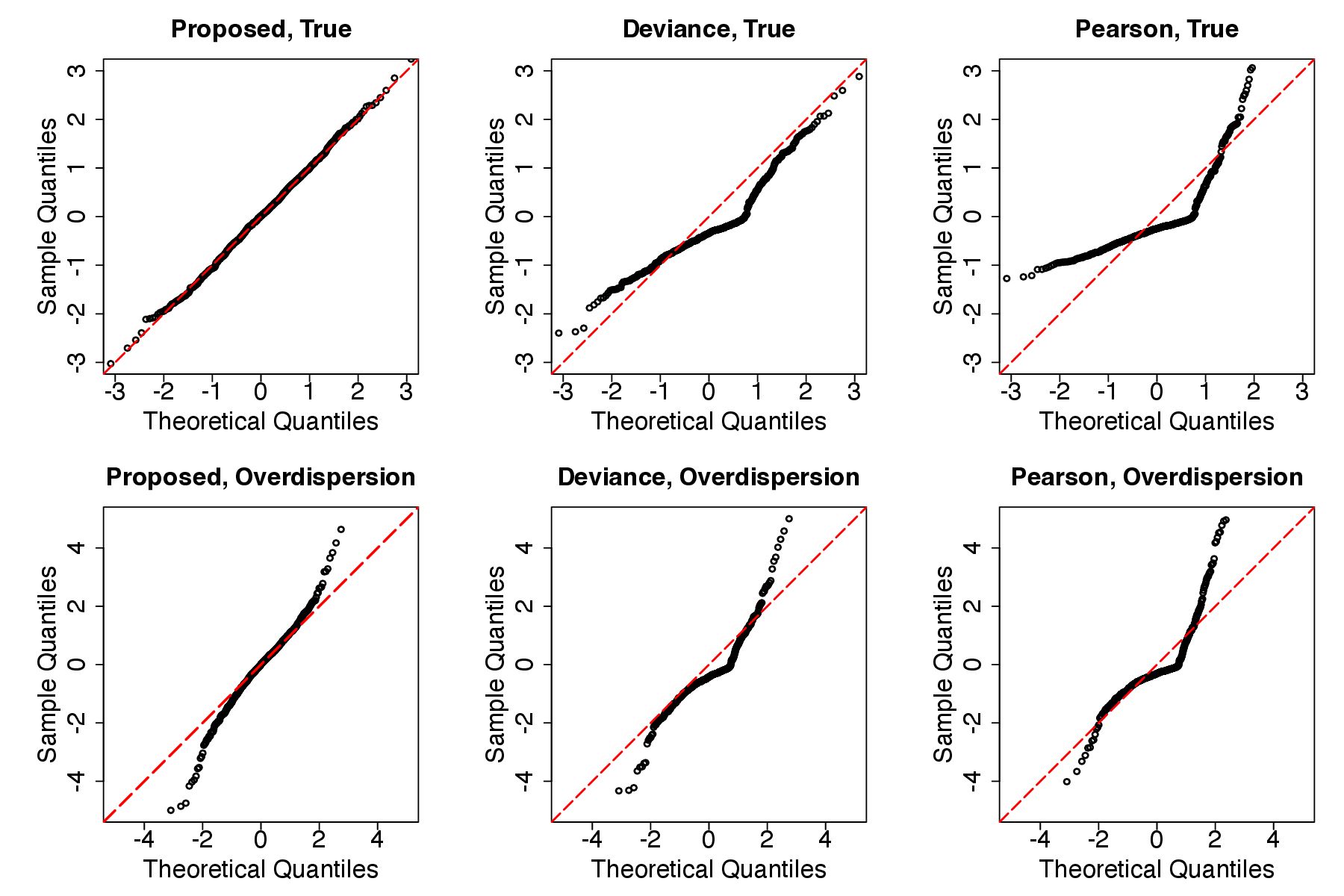} 
	
	\caption{QQ plots of the proposed residuals (left column)  compared with deviance (middle column) and  Pearson residuals (right column) for negative binomial data. The model is correctly specified in the first row, while in the second row, the data are mistakenly fit with a Poisson GLM.  \label{fig:over}}
\end{figure}

	For the negative binomial data described in Section \ref{sec:correct}, now we fit them with a Poisson GLM, and thus overdispersion is present. The resulting QQ plots of the proposed residuals, along with other residuals, are shown 
	 in the bottom row of Figure \ref{fig:over}. 
	In the first  column, we can see that the proposed residuals show a transition from being close to the diagonal when the model is correctly specified 
	to displaying  an obvious and readily interpretable discrepancy when overdispersion is an issue. In contrast, the  deviance and Pearson residuals show a large discrepancy with the null pattern in both scenarios,  making them not informative.

	In Appendix \ref{sec:unravel}, we  
	illustrate that the shape of the QQ plot for the proposed residuals is determined by the  relationship between the estimated distribution function  $\hat{F}_M$ and the true distribution function $F$. 
	Due to the orthogonality between the mean and dispersion components in GLMs, the mean structure is close to being correctly fit even when the dispersion parameter is misspecified.
	In an overdispersed model, due 
	to the underestimated  variance, $\hat{F}_M$ behaves more wildly than $F$. 
	Therefore, if one spots the {S-shaped pattern} in the QQ plot, as shown in the lower left panel of Figure \ref{fig:over}, it is probable that overdispersion is an issue.
%
%
%
%
%
%

\begin{figure}[!h]
	\centering
	\includegraphics[width=.9\textwidth]{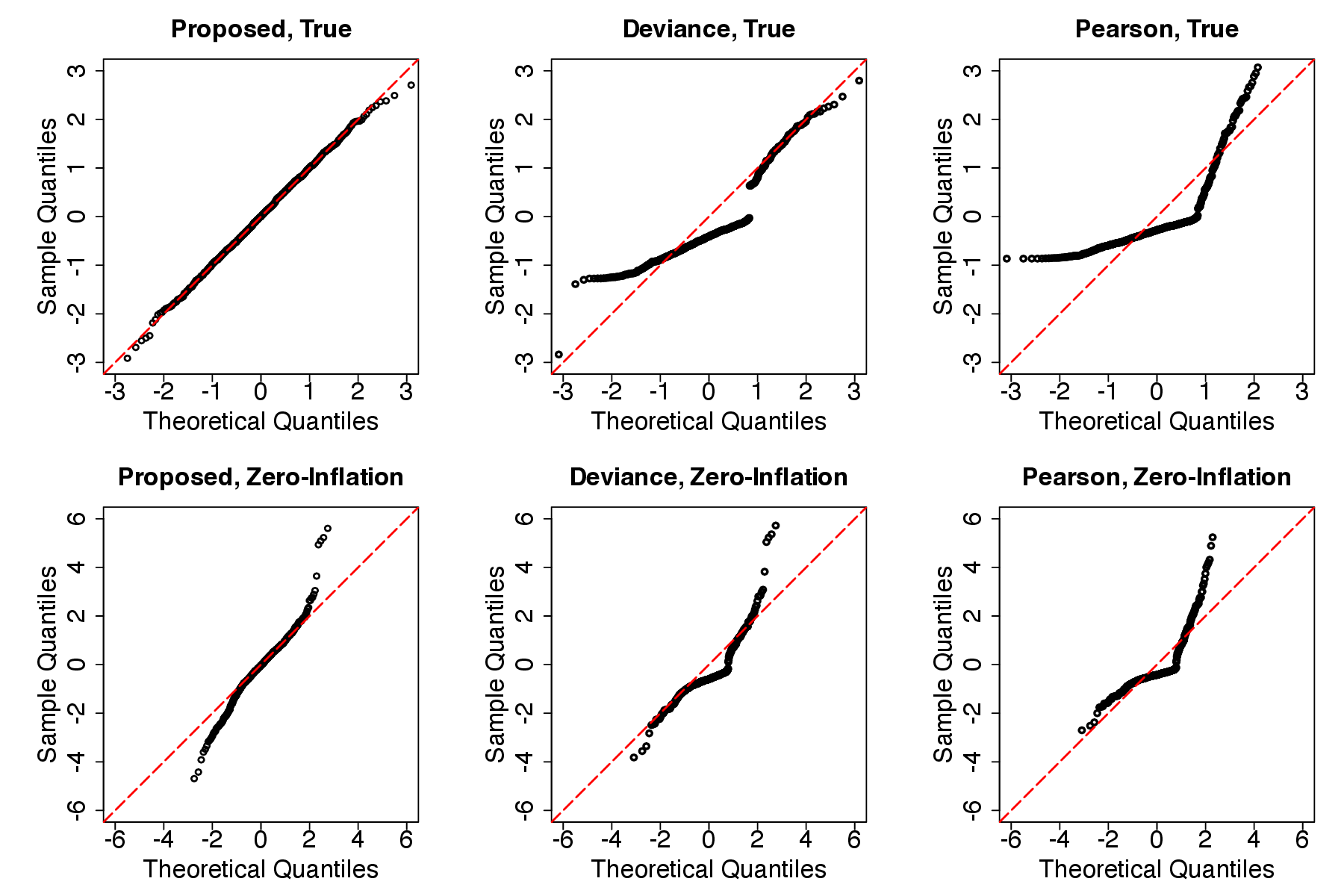} 
	
	\caption{QQ plots of the proposed residuals (left column) compared with the deviance (middle column)  the Pearson  residuals (right column) for zero-inflated Poisson data. The model is correctly specified in the top row and misspecified as a Poisson GLM in the bottom row. \label{fig:zero}}
\end{figure}

Besides negative binomial distributions, zero-inflated models are commonly used to handle overdispersion in count data. In Figure \ref{fig:zero}, we simulate  data using a zero-inflated Poisson model. The probability of excess
zeros is modeled with $\mathrm{logit}(p_0) = \beta_{00} + \beta_{10}X_1$, and the Poisson component has a mean $\lambda = \exp(\beta_0 +\beta_1X_1 +\beta_2X_2)$, where $X_1 \sim N(0,1)$ and $X_2$ is a dummy variable with a probability of 1 equal to 0.7, and $( \beta_{00} ,\beta_{10}, \beta_0, \beta_1, \beta_2) = (-2, 2, -2, 2, 1)$. 
In the top row of Figure \ref{fig:zero}, when the model is correctly specified, the proposed residuals closely follow their null pattern. In contrast, the deviance and Pearson residuals again deviate from normality. 
Here the saturated Poisson model is used to define the deviance residuals for the zero-inflated Poisson model (\citealt{lee2001analysis}).
When the model is misspecified,  as depicted in the bottom row of Figure \ref{fig:zero}, the proposed residuals show a large discrepancy from the diagonal. 
Moreover, a noticeable S-shaped pattern, similar to what was observed in the bottom left panel of Figure \ref{fig:over}, arises due to overdispersion.

\subsubsection{Ordinal Data}
		In this experiment, we consider ordinal regression models with three levels 0, 1, and 2. Under an ordinal logistic regression model with proportionality assumption, $P(Y\leq j)=F(\alpha_j),$  where $F$ is the distribution function of a  logistic random variable with mean $\beta_1X_1$. We let $\alpha_0=1$, $\alpha_1=4$, $\beta_1=3$, and $X_1\sim N(2,1)$. 
		
\begin{figure}[!h]\centering	\includegraphics[width=.9\textwidth]{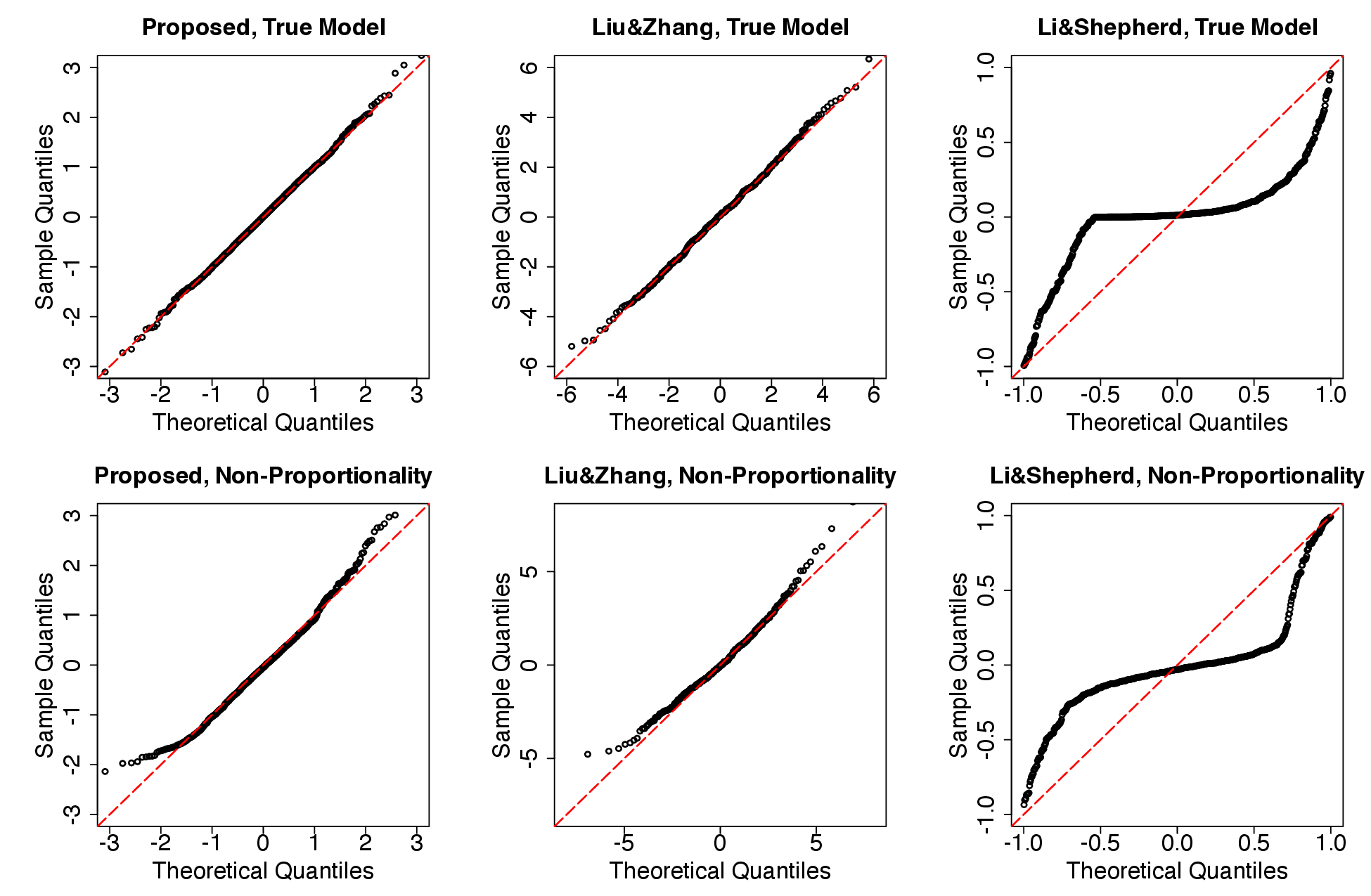}
			
			\caption{Graphical detection of non-proportionality in ordinal regression models.  The two rows correspond to  scenarios in which the assumption of proportionality is met and not met, respectively. \label{fig:prop}}
		\end{figure}

		Here we compare the proposed method with  
		the two types of residuals introduced in  \cite{li2012new} and \cite{liu2018residuals}   which were designed to tackle ordinal regression model diagnostics. 
		In Liu \& Zhang framework, it is assumed that there is a latent variable $A_i$ which follows a logistic distribution with mean  $X_{i1}\hat{\bm\beta}_1$. Given the fitted thresholds $\hat{\alpha}_0$ and $\hat{\alpha}_1$, they simulate the residual $\hat{r}_i^\text{Liu}$ from the  distribution  of $A_i|Y_i=y_i$. 
		Under the correct model, these residuals are expected to  follow a logistic distribution. 
		The Li \& Shepherd residuals are defined as $\hat{r}_{i}^\text{Li}=\hat{F}( Y_i-1|\mathbf{X}_i)+\hat{F}(Y_i|\mathbf{X}_i)-1$.  
		For continuous outcomes, these residuals are supposed to follow a uniform distribution over $[-1,1]$ under the true model.
			We can see  in the top row of Figure \ref{fig:prop} that
		both the proposed method and  the Liu \& Zhang residuals result in plots that are closely aligned with the diagonal when the model is correctly specified. 
		
		We then generate data under the scenario where the assumption of proportionality is not met, which is a common issue for ordinal regression models. Specifically,
		$P(Y\leq 0)=F(\alpha_0)$ as described above whereas $P(Y\leq 1)=F_1(\alpha_1)$, where   $F_1$ is the distribution function of a logistic random variable with mean $\beta_2X_1$ and we set $\beta_2=1$. The data are incorrectly fit with a proportional odds model, and the bottom row of Figure \ref{fig:prop} includes the results. 
	It is apparent  that our proposed method demonstrates  sensitivity to the presence of non-proportionality in this example. 

			To explore the shape of the QQ plot, in the supplementary material, we compare  the  underlying and fitted probabilities under non-proportionality. 
		The fitted probabilities of zeros and the fitted cumulative probabilities at $k_{\max}-1$   are consistently larger than the corresponding underlying probabilities, resulting in a QQ plot with both tails above the diagonal, as shown in the lower left panel of  Figure \ref{fig:prop}.

%

%
	
%

\subsubsection{Binary}\label{sec:binary}
\begin{figure}[!h]
	\centering
	\includegraphics[width=.9\textwidth]{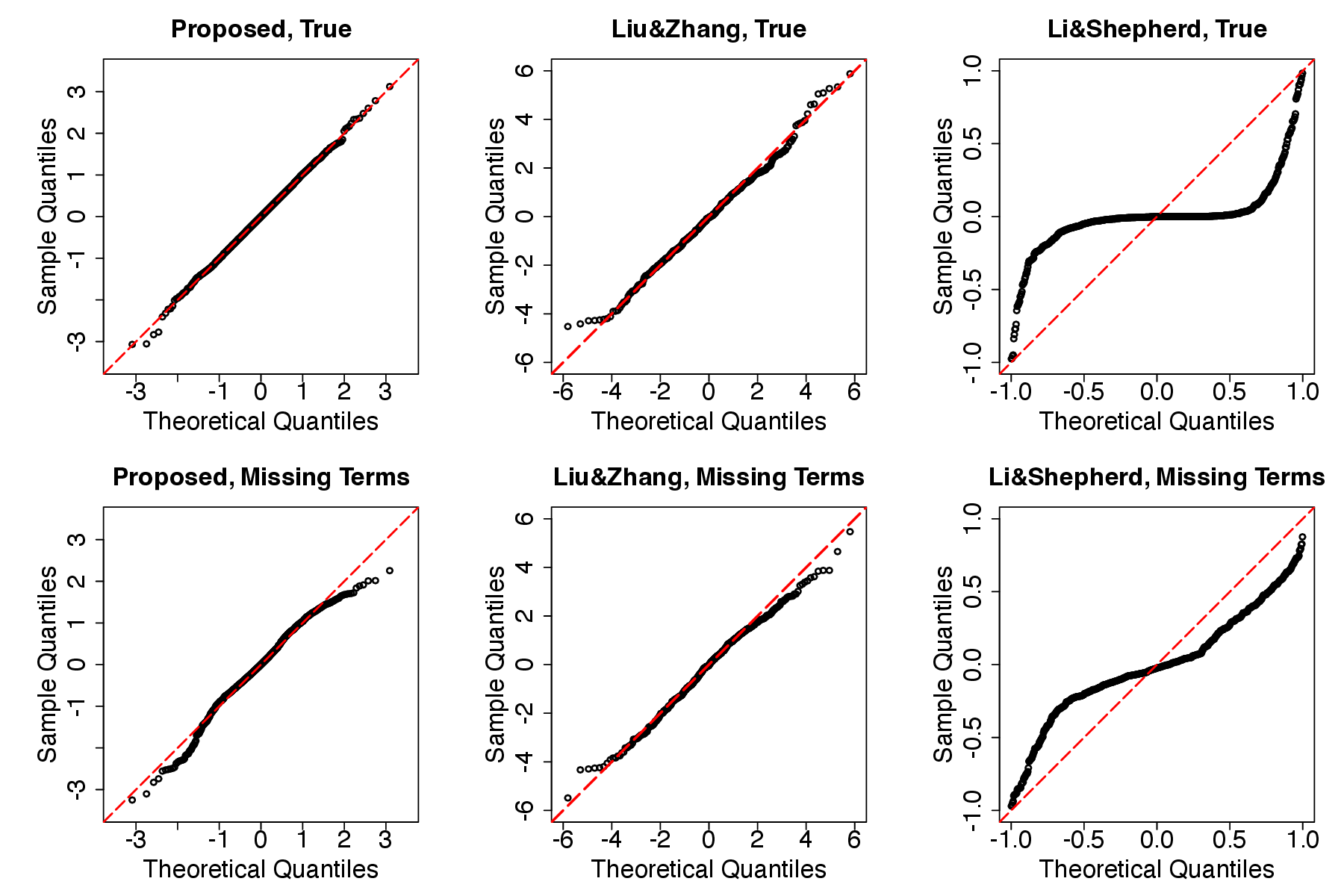} 
	
	\caption{QQ plots of the proposed residuals (left column) compared with  the Liu \& Zhang  residuals (middle column) and Li \& Shepherd residuals (right column) for the binary example. The model is correctly specified in the top row and a main effect and an  interaction term are missing  in the bottom row. \label{fig:binary3}}
\end{figure}

We further include an example of binary data. 
The underlying model is a logistic regression with the probability of 1 as $\mathrm{logit}^{-1}(\beta_0+\beta_1 X_1+\beta_2 X_2+\beta_3X_1 X_2)$, where $(\beta_0,\beta_1,\beta_2,\beta_3)=(-5,2,1,3)$, $X_1\sim N(1,1)$, and $X_2$ is a dummy variable with a probability of one equal to 0.7. For the misspecified model, the binary covariate and the interaction term are omitted. It was discussed in \cite{hosmer1997comparison} that most goodness-of-fit tests  tend to have limited power in such settings. 
Figure \ref{fig:binary3} shows the results. Due to the undesirable properties of deviance and Pearson residuals, here we display Liu \& Zhang and Li \& Shepherd residuals instead.
When the  model is correctly specified,  the proposed  residuals follow the null pattern.
When the interaction term is omitted, on the other hand,
{the proposed tool can effectively detect the model deficiency}. 

\subsubsection{Outliers}\label{sec:outlier}
\begin{figure}[!h]\centering	\includegraphics[width=.6\textwidth]{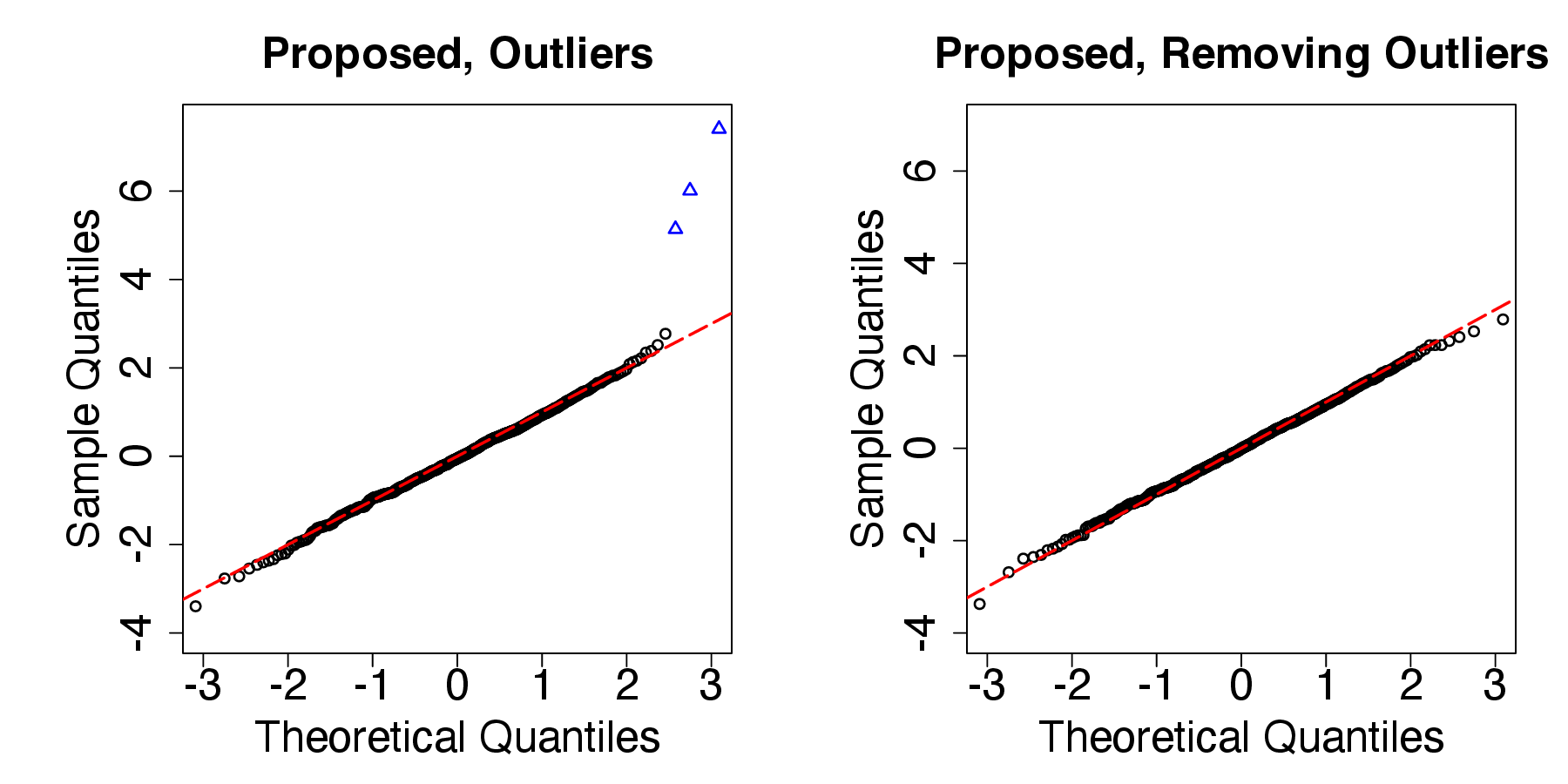}
	
	\caption{Graphical detection of outliers. \label{fig:out}}
\end{figure}

The proposed residuals can help identify outliers.  Figure \ref{fig:out} includes a Poisson example, with the same  underlying mean structure   as the negative binomial outcomes discussed in Section \ref{sec:correct}. We manually enlarged three outcomes by adding  values of $10,~15,$ and $20$ to them, respectively. In the left panel of Figure \ref{fig:out}, we can see the three modified data points stand out, signaling they are potential outliers.

We also emphasize that  close examination of the data is required for outlier identification.
As will be demonstrated in the next section, the distribution of the proposed residuals depends on the values of  covariates. Therefore, it is possible that a large value of the residual is the consequence of high leverage. 
When encountering suspected cases, it is important to carefully examine the data.

		\subsubsection{Limitation: Uninformative Residuals Versus Predictor Plots}
\begin{figure}[!h]\centering	\includegraphics[width=.6\textwidth]{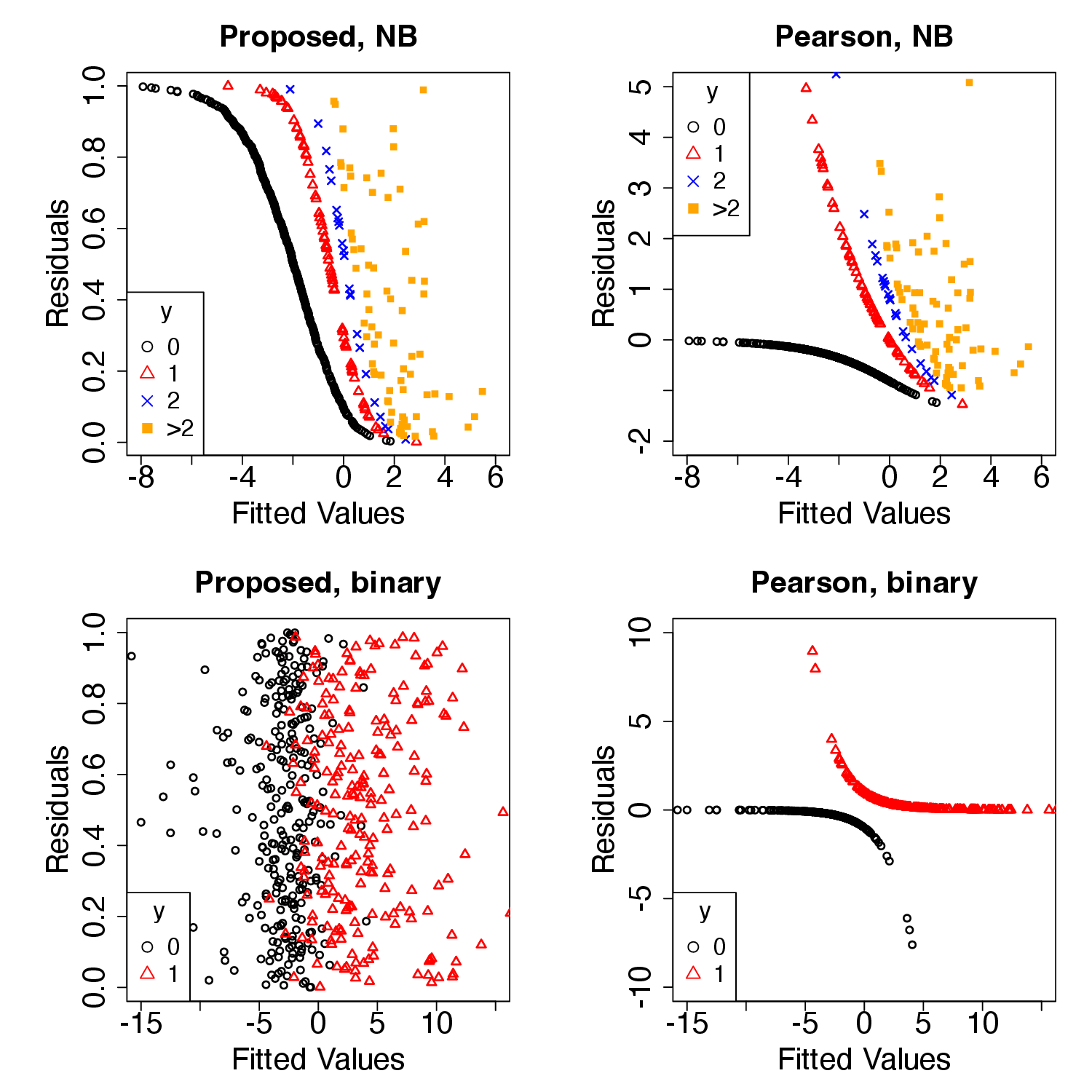}
			
			\caption{Residual versus fitted value plots when the model is correctly specified. \label{fig:versus}}
		\end{figure}

		 In linear regression models, another common utility of residuals is to check the mean structure. Analysts routinely check residuals versus predictor plots to decide whether the covariate structure is sufficient or if another variable should be included in the model. 
As discussed in Section \ref{sec:meancurve}, the distribution of the proposed residuals depends on the value of   covariates.
		In the left column of Figure \ref{fig:versus}, we display the residuals versus fitted values plots for the  negative binomial (NB) and binary examples of Sections \ref{sec:count} and \ref{sec:binary}, respectively, when the model is correctly specified.
		The fitted values are on the scale of the linear predictors.
		It is clear that the distribution of the residuals changes  with the fitted values, and we should not expect the
		residuals to exhibit  no discernible pattern. 
		In addition, we differentiate the residuals based on the corresponding outcomes. We can see that the residuals 
		show  lines of points corresponding to the observed responses, as noted in \cite{faraway2016extending}. 
		 It was shown  in the literature that other residuals, including Pearson,  deviance and Li \& Shepherd residuals, 
		 also face this challenge for discrete data (\citet{shepherd2016probability,liu2018residuals,liu2021assessing}), which we illustrate in the right column of Figure \ref{fig:versus}.
		In order to assess the mean structure,
		we proposed a tool in Section \ref{sec:meancurve}. We will demonstrate its utility in the next section.

		\subsection{Ordered Curve}\label{sec:simmean}
\begin{figure}[!h]
				\centering
				\includegraphics[width=.9\textwidth]{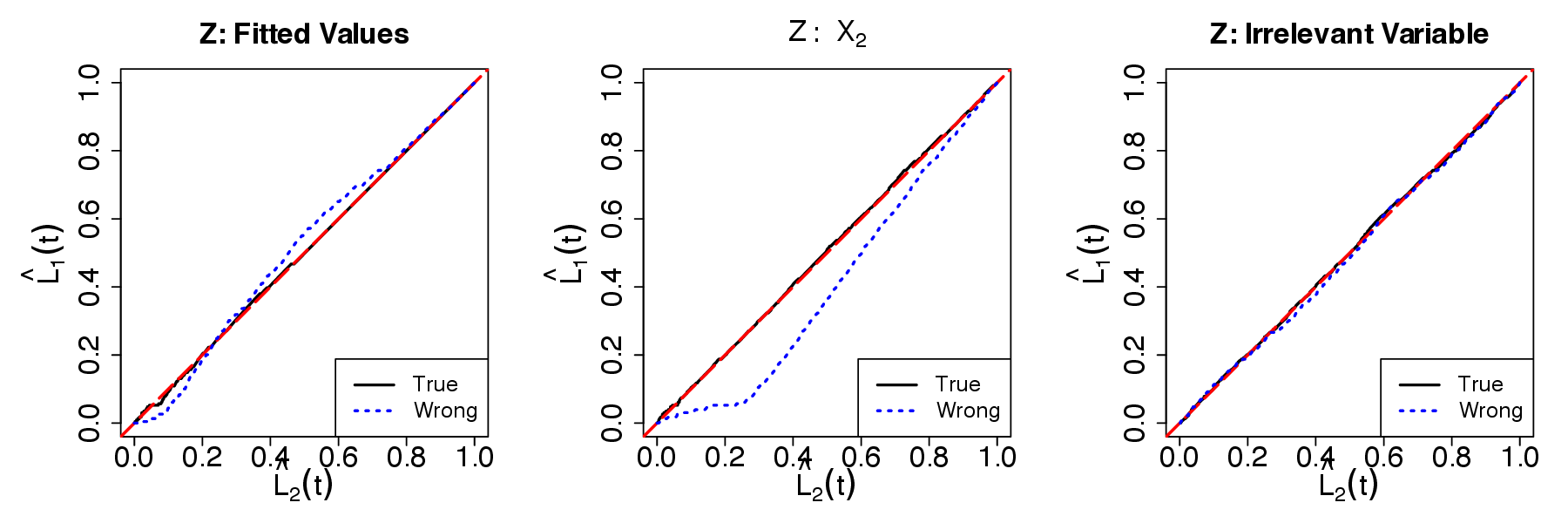} 
				
				\caption{
					Ordered curves of binary regression with different threshold variables $Z$. The sample size is 500.  \label{fig:meanbinary}}
			\end{figure}

		We first revisit the binary example in Figure \ref{fig:binary3}. In each plot of Figure \ref{fig:meanbinary}, we display the ordered curves for both the correctly specified (solid curve) and the misspecified model  (dotted curve). We consider three different choices for the threshold variable $Z$, the fitted values (left panel), the missing covariate (middle panel), and a randomly simulated irrelevant variable (right panel).
		We can see that, first,  regardless of the threshold variable, when the mean structure is correctly specified, the solid curves are close to the diagonal (dashed line). 
		Second, for the misspecified model, the degree of deviation between the ordered curve  and the diagonal depends on the choice of the threshold variable.
		When we use the fitted values to sort the data, the curve for the incorrect model   shows a small deviation from the diagonal.
		Strikingly, when the omitted variable $X_2$ is used as the threshold variable, it leads to  a curve far  from the diagonal. On the other hand, if $Z$ is a completely irrelevant variable, we can hardly detect the incorrectness in the mean structure. 
		
		We further explore a Poisson example. The mean function is $\mu =  \exp\left(\beta_0 +X_1\beta_1+X_2\beta_2\right)$, where $X_1$ and $X_2 \sim N(0,1)$ independently. The coefficients  are set to be $\beta_0 = 0, \beta_1 = 2$, and  $\beta_2=1$. 
		For the misspecified model, $X_2$ is omitted. Consistent with our observations of Figure \ref{fig:meanbinary}, using the omitted variable as the threshold variable elucidates the 
	distinction between the fitted means and the means of the actual outcomes.
	When we use the fitted values and an irrelevant variable as the threshold variable,
	we observe a slight deviation from the diagonal  under the misspecified model in this example. 
			\begin{figure}[!h]
			\centering
			\includegraphics[width=.9\textwidth]{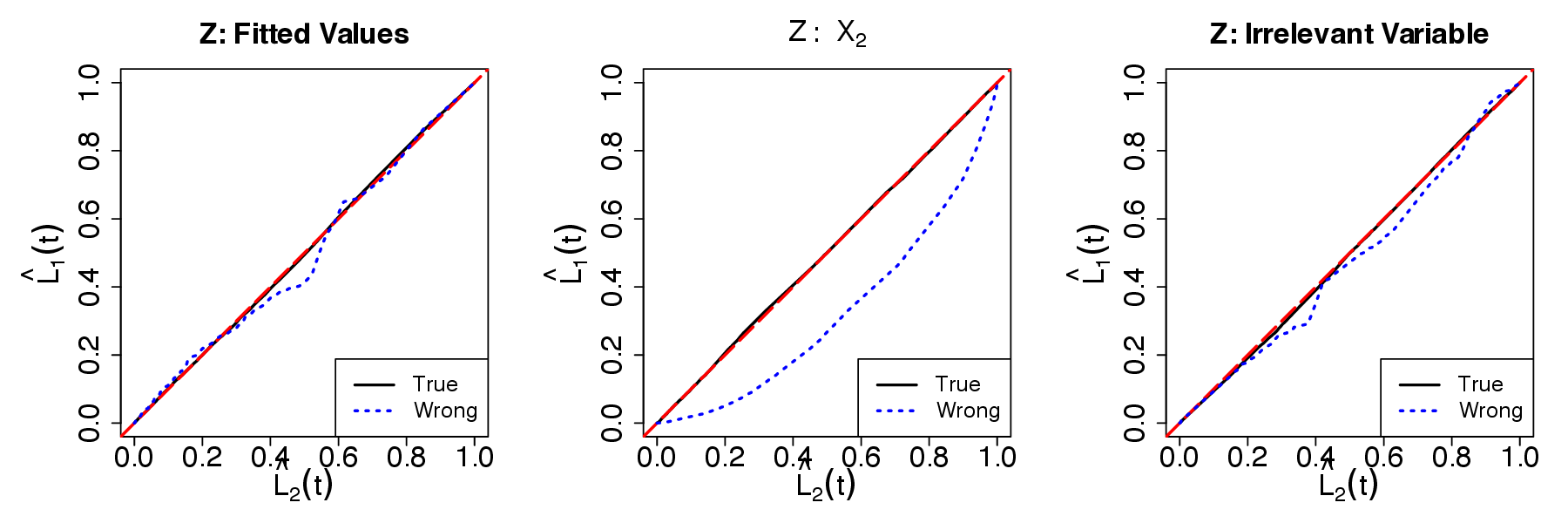} 
			
			\caption{
					Ordered curves of Poisson regression with different threshold variables $Z$. The sample size is 500. \label{fig:meanpois}}
		\end{figure}

\section{Data Analysis}\label{sec:data}
In this section,  we present real data examples to demonstrate the workflow of model assessment  using our tools.

\subsection{Count Data}

AT\&T ran an experiment varying five factors relevant to a wave-soldering procedure for mounting
components on printed circuit boards
(\citealt{comizzoli1990robust}). The response variable is the count of how many solder
skips appeared to a visual inspection.  Table \ref{tab:covariates} includes the description and summary statistics of the  important covariates.
The sample size is 900. 
This example demonstrates the application of the proposed residuals even in cases where continuous covariates are not present, as long as there are factors with many levels.

\begin{table}[!h]\centering
	\caption{Covariates of AT\&T data.\label{tab:covariates}}
	\begin{tabular}{clll}
		\toprule
Variable&Description&Levels (Frequency)\\
\midrule
Opening&the amount of clearance around the&L (300), M (300), S (300) \\& mounting pad\\
\midrule
Solder&the amount of solder&Thick (450), Thin (450)\\
\midrule
Mask& type and thickness of the material &A1.5 (180), A3 (270), \\&used for the solder mask& A6 (90), B3 (180), B6 (180)\\
\midrule
PadType&the geometry and size of the&D4 (90), D6 (90), D7 (90),\\&  mounting pad&L4 (90), L6 (90), L7 (90),  \\&&L8 (90), L9 (90), W4 (90), \\&&W9 (90)\\\midrule
Panel&each board was divided into 3 panels&1 (300), 2 (300), 3 (300)&\\
\bottomrule
	\end{tabular}

\end{table}

\begin{figure}[!h]
	\centering
	\includegraphics[width=.9\textwidth]{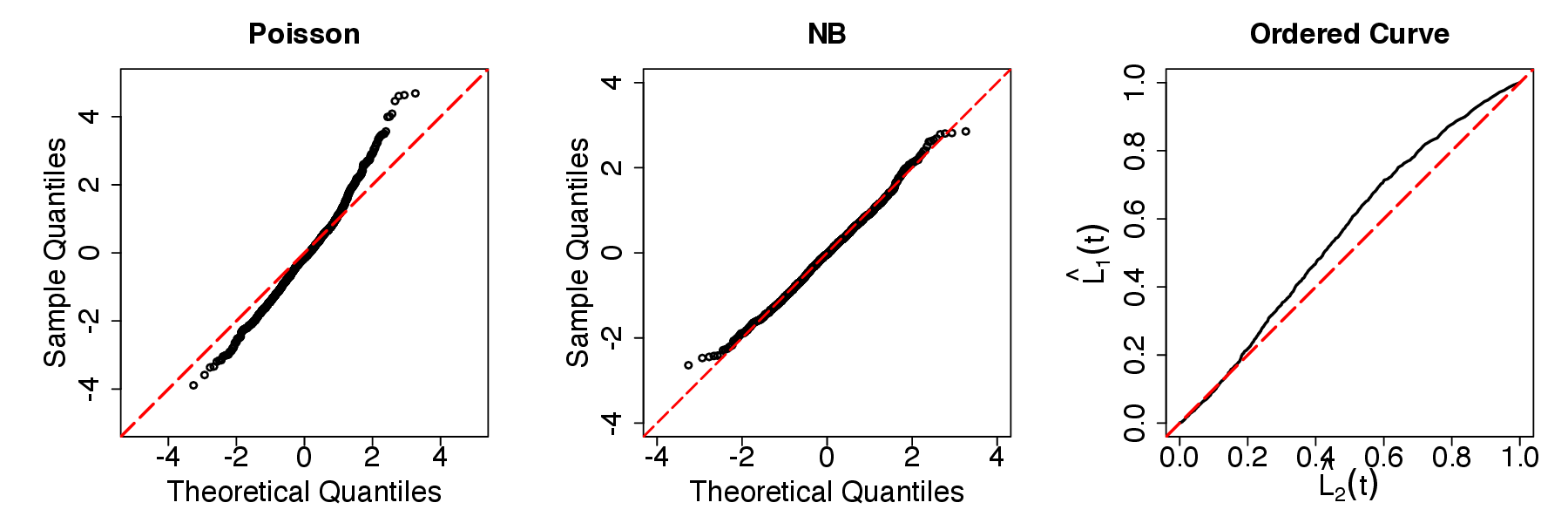} 
	
	\caption{
		Assessment plots for the regression models of the solder data with main effects only. The left and middle panels display the proposed residuals of  Poisson and negative binomial regressions, respectively. The right panel shows the ordered curve and the fitted values are used as the threshold variable. \label{fig:solderpois}}
\end{figure}

We first fit  regression models with main effects.
In the left panel of  Figure \ref{fig:solderpois},  we use a Poisson GLM, and
 our tool displays a pronounced S-shaped pattern, hinting at potential overdispersion. 
 This is consistent with the conclusion of \cite{faraway2016extending}. In the middle panel, we use a negative binomial  distribution to correct for overdispersion. According to our residuals, the negative binomial distribution  leads to a substantial improvement in the model fit, yet some insufficiency is indicated by the tails.
 In the right panel, we analyze the ordered curve, which implies insufficiency in the mean structure.

\begin{figure}[!h]
	\centering
	\includegraphics[width=.6\textwidth]{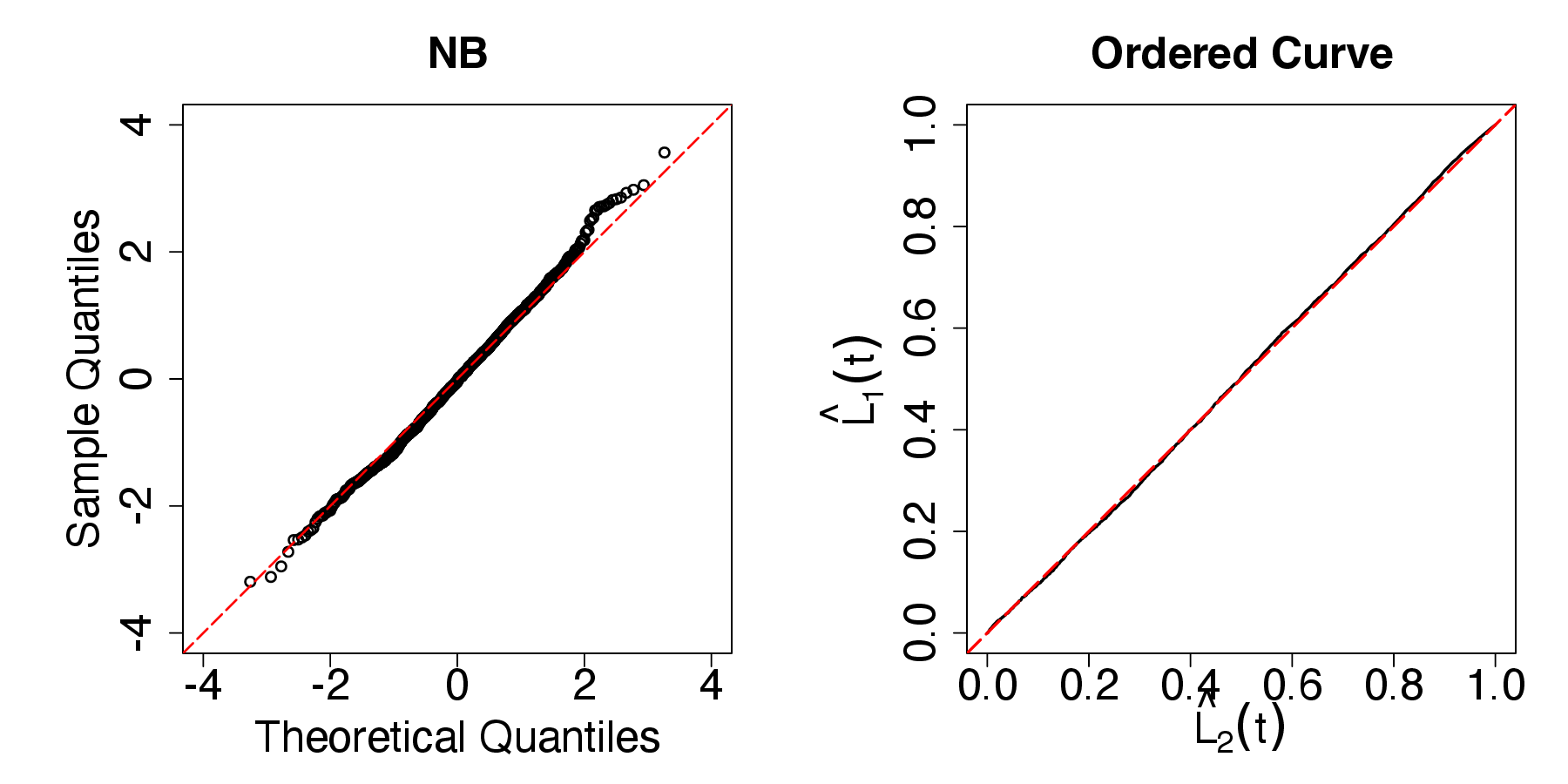} 
	
	\caption{
		Assessment plots for the negative binomial models of the solder data with main effects and interactions.  \label{fig:soldernb}}
\end{figure}

We then include interaction terms  selected using a stepwise procedure.
Figure 
\ref{fig:soldernb} shows the diagnostic plots of the resulting model. The inclusion of   interaction terms yields a considerable improvement in model fit.

\subsection{Binary Outcomes}

 The dataset was collected in 
one of the earliest studies addressing the factors impacting the chance of developing heart disease (\citealt{rosenman1975coronary}). The study started in 1960 and involved 3154 healthy men aged between 39 and 59. All the subjects were free of heart disease  at the beginning of the study. The response variable is  whether these men  developed  heart disease 
eight and a half years later. 255 men developed coronary heart disease,  while the others did not. 
Table \ref{tab:covariatesheart} includes the variables that might be related to the chance of developing this disease.

\begin{table}[!h]\centering
		\caption{Covariates of heart data.\label{tab:covariatesheart}}
	\begin{tabular}{cll}
		\toprule
	Variable&Description&Mean (sd) / Levels (count)\\
	\midrule
	age&
	age in years&46.275 (5.517)\\
	\midrule
	height&
	height in inches&69.780 (2.521)\\
		\midrule
	sdp&
	systolic blood pressure in mm Hg&128.603 (15.056)\\
		\midrule
	chol&
	fasting serum cholesterol in mm \%&226.346 (43.421)\\
		\midrule
	behave&
	behavior type which is a factor&A1 (263),  A2 (1320), B3 (1209), \\&& B4 (348)\\
		\midrule
	cigs&
	number of cigarettes smoked per day&11.577 (14.494)\\
		\midrule
%
	arcus&
	arcus senilis&  absent (2202), present (938)\\
		\midrule
	bmi&
	body mass index & 24.516 (2.564) \\
	\bottomrule
	\end{tabular}
\end{table}

\begin{figure}[!h]
	\centering
	\includegraphics[width=.6\textwidth]{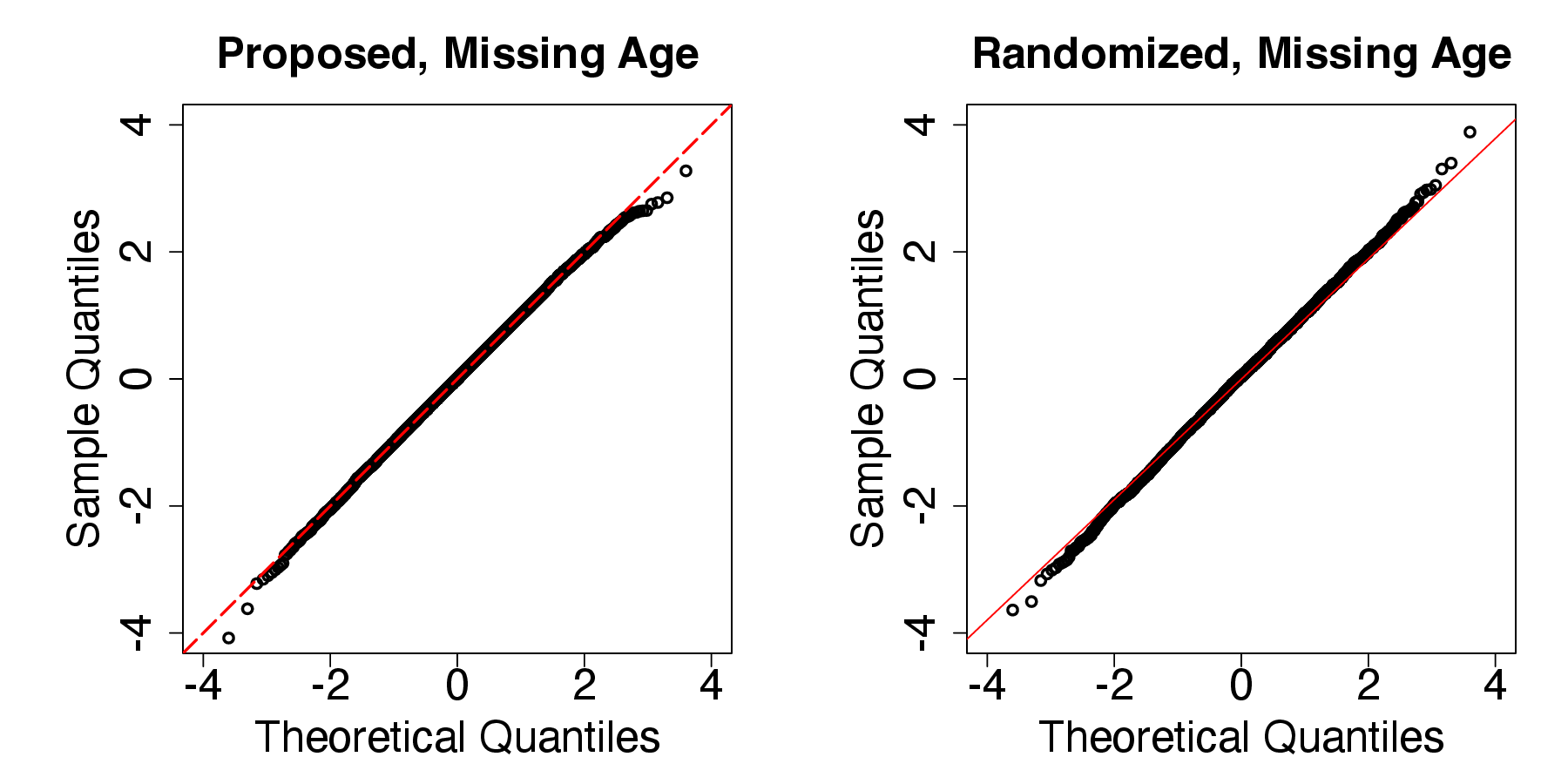} 
	
	\caption{Residual plots for logistic regression of  the heart disease data with the variable age missing. Left: QQ plot of the proposed residuals. Right: QQ plot of  randomized quantile residuals. 
		\label{fig:wsgmmiss}}
\end{figure}

We first highlight the difficulties in the  assessment of regression models with binary outcomes.  We fit the data with all the variables except age, which is a very important predictor clinically. Figure \ref{fig:wsgmmiss} includes the proposed residual plot (left panel) and the randomized quantile residual plot (right panel). The proposed residuals have a slight deviation from the diagonal at the upper tail, while the randomized quantile residuals  indicate that the model is  sufficient.  This is further supported  by the Hosmer-Lemeshow  test (\citealt{hosmer1980goodness}), which returns a large p-value 0.62, suggesting the adequacy of the model. 
Since the data are fit with logistic regression, a calibration plot is not revealing either.

\begin{figure}[!h]
	\centering
	\includegraphics[width=.6\textwidth]{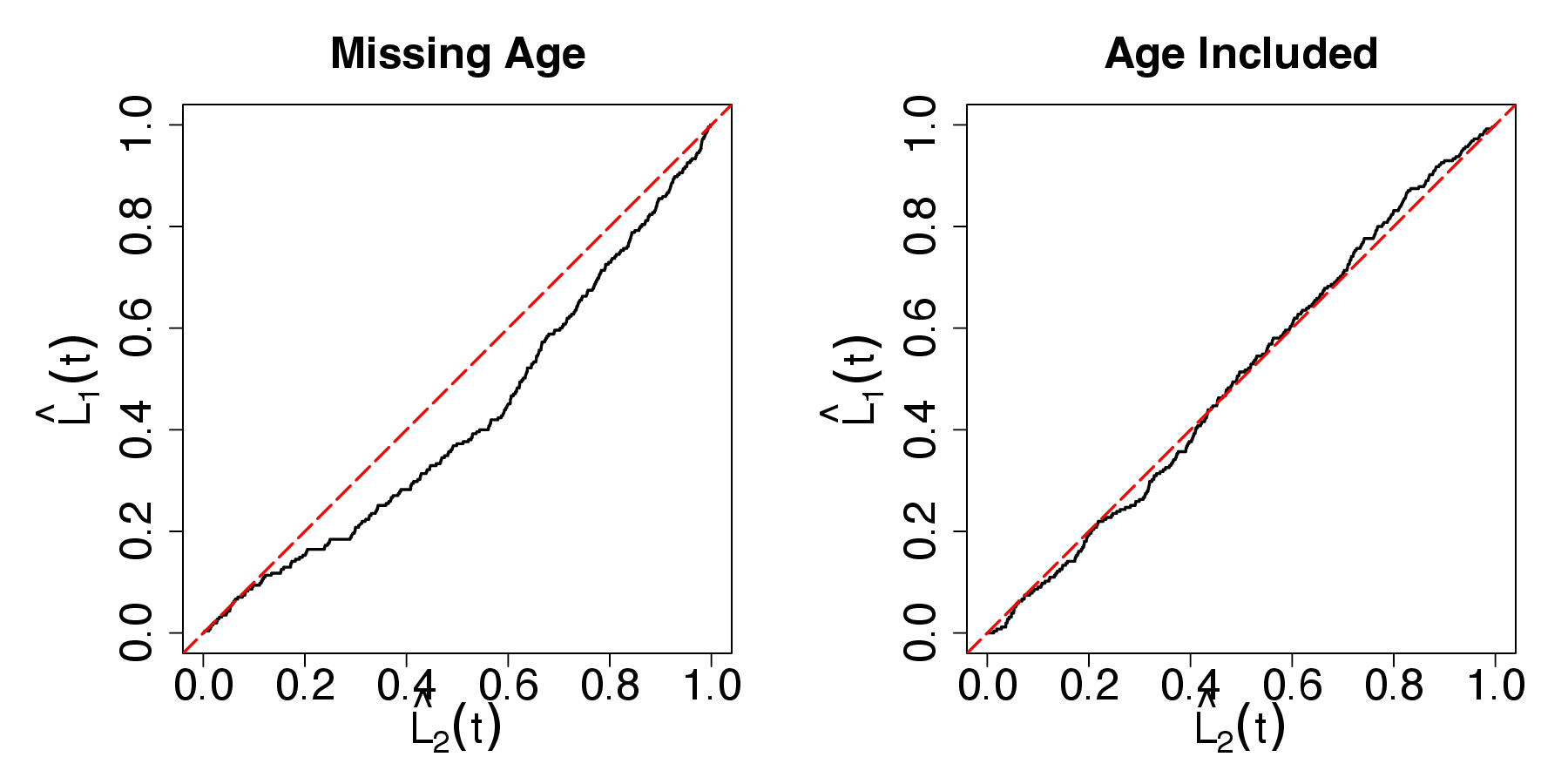} 
	
	\caption{Ordered curves for logistic regression of  the heart disease data. Left: age is not included in the mean and age is used as the threshold variable. Right: age is included and fitted values are used as the threshold variable. \label{fig:wsgm}}
\end{figure}

However, in the left panel of Figure \ref{fig:wsgm}, we show the ordered curve when age  is used as the threshold variable. The curve clearly suggests that the mean structure is insufficient and age should be considered as a predictor, which is consistent with common knowledge.
In the right panel of Figure \ref{fig:wsgm}, we include age as a predictor, and  the ordered curve suggests an improvement in the mean structure.

\section{Conclusion}\label{sec:conc}
Regression models with discrete outcomes are commonly used in a wide range of areas. Diagnostics for such  models are challenging due to the lack of effective tools. In this paper, we proposed the DPIT residuals to assess regression models with discrete outcomes. 
To further assess the mean structure, we also proposed an ordered curve, and we showed in the data analysis that it can reveal model deficiency  overlooked by other tools.  
We focused on diagnostics and informal assessment in this paper. Goodness-of-fit tests (e.g., \citealt{hosmer1997comparison,nattino2020assessing}) which provide p-values and statements with statistical confidence will be investigated in the future.
	
To summarize the workflow of model assessment using the proposed tools,
one should first look into the QQ plot of the DPIT residuals for overall assessment.  
Meanwhile, 
one should  examine the ordered curve to evaluate the mean structure. 
We recommend  using the fitted values and  potential predictors as the threshold variable. If one variable leads to a large discrepancy between the ordered curve and the diagonal, this variable should be considered to include in the mean function. On the other hand,
if the mean structure seems sufficient yet the residual plot indicates model deficiencies, one can identify causes of misspecification from the shape of the QQ plot. For instance, an S-shaped QQ plot for count data might hint at overdispersion, while a U-shaped plot for ordinal outcomes might imply non-proportionality. This can also be combined with tools devoted to specific types of misspecification (e.g., \cite{pregibon1981logistic}).
We note that in practice,  we might incur more than one type of misspecification simultaneously, and the causes of misspecification are not always  identifiable (\citealt{cook1999graphs}). 
The regression model-building process should therefore involve iterations between assessment  and improvement.

	\appendix
			\section{Additional Derivations}\label{sec:add}

				Under the true model, the following  properties  hold   for the DPIT.
			\begin{enumerate}
				\item 
				$\mathrm{E}\left\lbrace{r}(Y|\mathbf{X})\right\rbrace=\frac{1}{2}.$ On the normal scale, $\mathrm{E}\left\lbrace\Phi^{-1}\left[{r}(Y|\mathbf{X})\right]\right\rbrace=0.$
				\item 
				$\mathrm{Var}\left\lbrace{r}(Y|\mathbf{X})\right\rbrace=\frac{1}{12}.$ On the normal scale, $\mathrm{Var}\left\lbrace\Phi^{-1}\left[{r}(Y|\mathbf{X})\right]\right\rbrace=1.$
				\item For $y_1\leq y_2$,
				$${r}(y_1|\mathbf{X})\leq {r}(y_2|\mathbf{X}).$$
			\end{enumerate}
			The third property holds since both $F(\cdot|\mathbf{X})$ and $G_0(\cdot)$  are  non-decreasing functions.
			With parameter estimates plugged in, the residuals resemble the DPIT, which we show in Section \ref{sec:asym}.

		 We now
			 elucidate the construction of $\hat{G}_{Mi}$ in \eqref{eq:empgm}.
			If we were to use all the observations, 
			the estimated DPIT of the $i$th outcome is
			\begin{align*}
				&\frac{1}{n}\sum_{j=1}^n {F}\left({F}^{(-1)}(\left.{F}(Y_{i}| \mathbf{X}_{i})\right| \mathbf{X}_{j})| \mathbf{X}_{j}\right)\\=&\frac{1}{n}\sum_{j=1,j\neq i}^n {F}\left(\left.{F}^{(-1)}({F}(Y_{i}| \mathbf{X}_{i})| \mathbf{X}_{j})\right| \mathbf{X}_{j}\right)+\frac{1}{n}{F}\left(\left.{F}^{(-1)}\left({F}(Y_{i}| \mathbf{X}_{i})| \mathbf{X}_{i}\right)\right| \mathbf{X}_{i}\right)\\=&\frac{1}{n}\sum_{j=1,j\neq i}^n {F}\left(\left.{F}^{(-1)}({F}(Y_{i}| \mathbf{X}_{i})| \mathbf{X}_{j})\right| \mathbf{X}_{j}\right)+\frac{{F}(Y_{i}| \mathbf{X}_{i})}{n}.
			\end{align*}
			The first term in this display resembles ${G}_{0}\left({F}(Y_{i}| \mathbf{X}_{i})\right)$,  the DPIT.
			The second term, however, 
			induces bias from ${G}_{0}\left({F}(Y_{i}| \mathbf{X}_{i})\right)$. 
			Hence, we ought to exclude the $i$th observation.
			
						Next, we derive the $L_2(t)$ function for ordered curves.
			If 
			$Z$ is contained in $\mathbf{X}$,
			the cumulative response equals  $$\mathrm{E}\left[Y1(Z\leq t)\right]=\mathrm{E}_\mathbf{X}\mathrm{E}\left[Y1(Z\leq t)|\mathbf{X}\right]
			=\mathrm{E}_\mathbf{X}\left[1(Z\leq t)\mathrm{E}\left(Y|\mathbf{X}\right)\right]
			=\mathrm{E}\left[\Lambda1(Z\leq t)\right].$$
			If $Z$ is independent of $\mathbf{X}$ and $Y$, 
			the cumulative response can be written as $$\mathrm{E}\left[Y1(Z\leq t)\right]=
			\mathrm{E}\left[1(Z\leq t)\right]\mathrm{E}\Lambda=\mathrm{E}\left[1(Z\leq t)\Lambda\right].$$
			In either case,  $$\mathrm{E}\left[Y1(Z\leq t)\right]={\mathrm{E}\left[\Lambda1(Z\leq t)\right]}.$$

		We establish Theorem \ref{main} with the following assumptions.
		\begin{assumption}\label{op}
			
			$\hat{  \bm\beta}$ is asymptotically efficient. The sequence $\sqrt{n}(\hat{  \bm\beta} -\bm\beta_0)$ converges in distribution to  a tight, Borel-measurable random element.
		\end{assumption}
		The  maximum likelihood estimator  of GLMs satisfies the  asymptotic efficiency assumption  under regularity conditions. 
		
		\begin{assumption}\label{bound}
			The density of ${G}_{\bm\beta }\left({F}(Y_i| \mathbf{X}_i,\bm\beta)\right)$ is bounded for $\bm\beta$ ranging over a small neighborhood of $\bm\beta_0$.
		\end{assumption}
		
		Assumption \ref{bound} holds if ${F}(k| \mathbf{X},\bm\beta)$ is Lipschitz continuous with respect to $\bm\beta$, for a fixed $k$. 
		 		We verify the assumptions for GLMs in the supplementary material.
		
		\begin{assumption}[Lipschitz condition]\label{lips}There exists a constant $\alpha_1$ such that for  $\bm\beta$ and $\bm\beta'$ in  a small neighborhood of $\bm\beta_0$,
			\begin{gather*}
				\left|	{G}_{\bm\beta }\left({F}(y| \mathbf{x},\bm\beta )\right)-{G}_{\bm\beta'}\left({F}(y| \mathbf{x},\bm\beta')\right)\right|
				\leq  \alpha_1\left|\bm\beta-\bm\beta'\right|.
			\end{gather*}
		\end{assumption}
		

\section{Additional Simulation}\label{sec:addsimu}
\begin{figure}[!h]
	\centering
	\includegraphics[width=.6\textwidth]{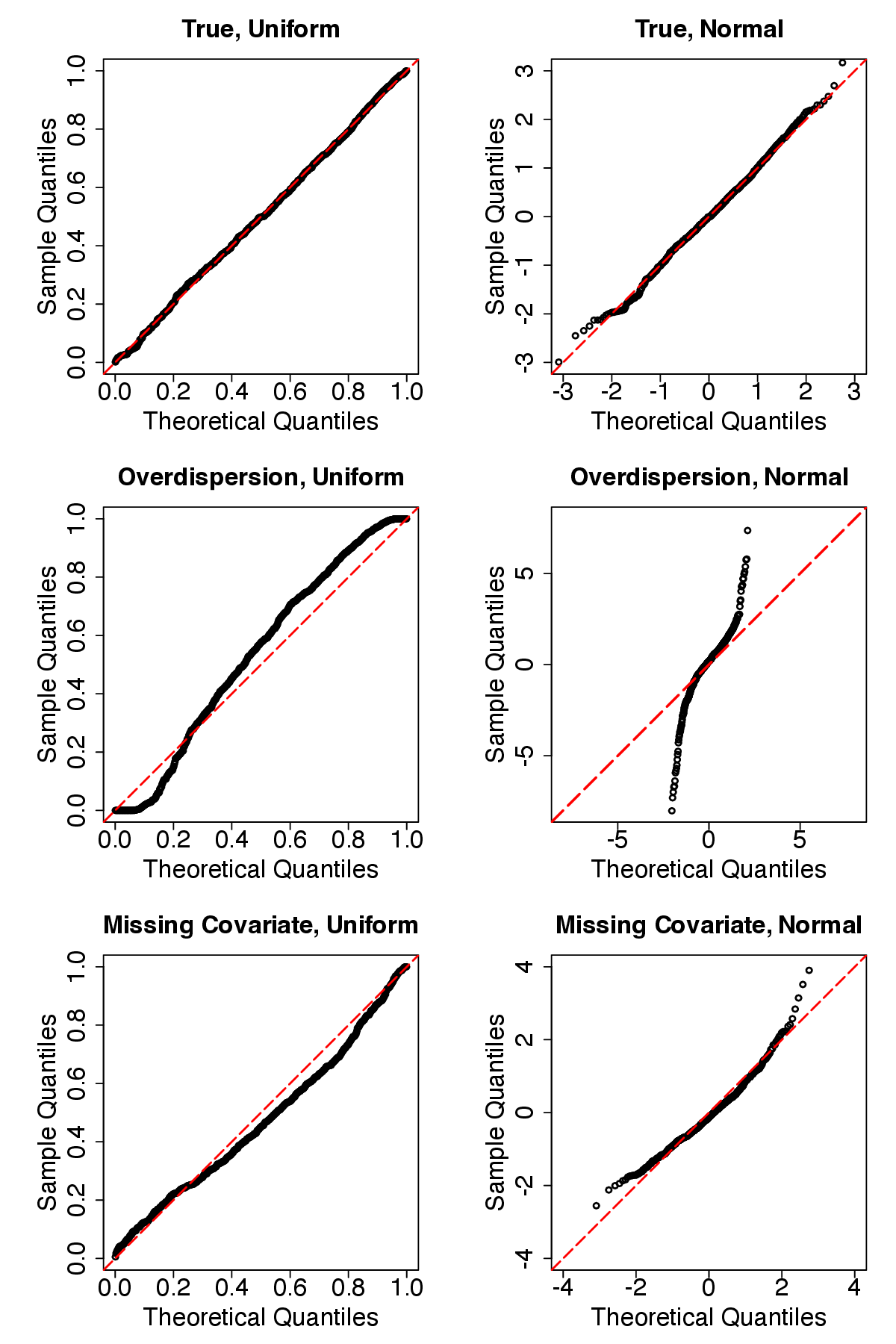} 
	\caption{QQ plots of the proposed residuals under the  uniform and normal scales. \label{fig:normal}}
\end{figure}

To compare the usage of our residuals on a normal and a uniform scale, in Figure \ref{fig:normal}, we display the QQ plots of the original residuals $\hat{r}(Y_i|\mathbf{X}_i)$ against the quantiles of a uniform distribution in the left column, and $\Phi^{-1}\left[\hat{r}(Y_i|\mathbf{X}_i)\right]$ against the quantiles of a standard normal distribution in the right column. 
We simulate  data using a negative binomial distribution, and the data are fit correctly in the top row. In the middle row, we fit the data with a Poisson GLM, while in the bottom row, one of the covariates is missing.
We can see in the middle and bottom rows that  the uniform residuals accentuate the center, while the normal residuals emphasize  the tails; see a thorough discussion in \cite{gan1991probability}.
In practice, one can choose either or both displays for model assessment.
Note that infinite values might occur for normal residuals, when the original residuals are very close to 0 or 1.

	\subsection{Unraveling QQ Plots}\label{sec:unravel}
When the model is correctly specified, our residuals should follow a uniform distribution. To construct QQ plots, one can plot the residuals against the quantiles of a uniform distribution.
To understand the shape of  the QQ plots of the  proposed residuals, here we  reproduce their patterns in a  heuristic manner.
We define the unobservable auxiliary residuals  $$	\hat{r}_{M0}(Y_i|\mathbf{X}_i)={G}_{M0}\left(\hat{F}_M(Y_{i}| \mathbf{X}_{i})\right),  i = 1, \ldots , n ,$$
where ${G}_{M0}=			\mathrm{E}_\mathbf{X}\left[F\left(\left. F_M^{(-1)}(s|\mathbf{X})\right|\mathbf{X}\right)\right]$ is the underlying distribution of ${F}_M(Y| \mathbf{X})$.
The auxiliary residuals  follow a uniform distribution. 
Since  ${G}_{M0}$  is  a monotone function, and  $\hat{G}_{Mi}$ converges to $\mathrm{E}_\mathbf{X}\left[F_M\left(\left. F^{(-1)}_M(s|\mathbf{X})\right|\mathbf{X}\right)\right]$, which is also a monotone function,
the rank of $	\hat{r}(Y_i|\mathbf{X}_i)$ among  the proposed residuals is approximately same as the rank of 
$	\hat{r}_{M0}(Y_i|\mathbf{X}_i)$ among the auxiliary residuals.
Therefore, the QQ plot of $	\hat{r}(Y_i|\mathbf{X}_i),i=1,\ldots,n$ could, in theory, be approximated by plotting $$	\hat{r}(Y_i|\mathbf{X}_i)\text{ against }	\hat{r}_{M0}(Y_i|\mathbf{X}_i),i=1,\ldots,n.$$
The shape of the QQ plot of  proposed residuals is thus determined by the  relationship between $\hat{G}_{Mi}$ and  $G_{M0}$. Furthermore, this relationship is preponderantly determined by the distinction between $\hat{F}_M$ and
$F$, averaging over covariates. Therefore, if the QQ plot is above the diagonal, it implies that $\hat{F}_M(k)>F(k)$ on average, and vice versa. 

This property of the proposed residuals  can help identify certain causes of misspecification. 
In particular,  the behavior in the lower tail displays the comparison between  $\hat{F}_M(k)$  and $F(k)$ for a small $k$. Conversely, the behavior in the  upper tail reflects the contrast between  $\hat{F}_M(k)$ and $F(k)$ for a large $k$.

\section*{Acknoledgements}

I thank Dr. James Hodges and the anonymous reviewers for their helpful comments. I gratefully acknowledge funding of the National Science Foundation (DMS-2210712).

\section*{Supplementary Material}


\begin{description}
		\item[Supplementary material:] The supplementary material includes additional simulation results and proofs of the theoretical results in Sections \ref{sec:simulation} and  \ref{sec:asym}. (.pdf file)
	\item[\texttt{R} code and package:] The \texttt{R} code for simulation and data analysis.
The proposed methodology is implemented in the \texttt{assessor}  package. 
	  See the README contained in the zip file for more details.
(code202312.zip, zip file)
\end{description}
	\singlespacing
\bibliography{pitt.bib}
\bibliographystyle{chicago}

		\end{document}